\documentclass[12pt,preprint]{aastex}

\newcommand{\aan}{$\alpha$($\alpha$n,$\gamma$)$^{9}$Be}
\newcommand{\atg}{$\alpha$(t,$\gamma$)$^7$Li}
\slugcomment{Submitted to Astrophysical J.}
\shorttitle{Collapsed Cores in Globular Clusters}
\shortauthors{Djorgovski et al.}
\begin{document}
\title{Sensitivity of r-Process Nucleosynthesis to Light-Element Nuclear Reactions}
\author{Taka. Sasaqui\altaffilmark{1,2,6}, T. Kajino\altaffilmark{1,2}}
\affil{National Astronomical Observatory of Japan}
\affil{Department of Astronomy, Graduate School of Science, University
of Tokyo, 7-3-1 Hongo, Bunkyo-ku, Tokyo 113-0033, Japan}
\email{sasaqui@th.nao.ac.jp}
\author{G. J. Mathews\altaffilmark{3}, K. Otsuki\altaffilmark{3,5}}
\affil{Center for Astrophysics, 
University of Notre Dame, Notre Dame,
IN 46556, U.S.A.}
\and

\author{T. Nakamura\altaffilmark{4}}
\affil{Department of Physics, Tokyo Institute of Technology,
2-12-1 O-Okayama Meguro-ku, Tokyo 152-0033, Japan}

\begin{abstract}
We study the efficiency and sensitivity of
r-process nucleosynthesis to 18  light-element nuclear reaction rates.
We adopt empirical power-law relations to parameterize the reaction
sensitivities.
We utilize two different hydrodynamic models 
for the neutrino-driven winds
in order to study the dependence of our result on 
supernova wind models. We also utilize an exponential model 
to approximate a wide variety of other
 plausible conditions for the r-process.
We identify  several specific nuclear
reactions among light neutron-rich nuclei that play a critical role in 
determining the final r-process nucleosynthesis yields.
As an illustration, we examine ``semi-waiting'' 
points among the carbon
isotopes.  We show that  not only neutron capture and 
 $\beta$-decay, but also $(\alpha, \mathrm{n})$ reactions 
 are important in determining waiting points along 
 the r-process path. 
Our numerical results from this sensitivity analysis 
 serve foremost to clarify which light nuclear 
 reactions are most influential
in determining the final r-process abundances.  We also 
quantify  
the effects of present nuclear uncertainties on the final r-process abundances.
This study thus emphasizes and
motivates which future determinations of nuclear reaction rates
will most strongly impact our understanding of 
r-process nucleosynthesis. 
\end{abstract}

\altaffiltext{1}{National Astronomical Observatory, 
2-21-1 Osawa, Mitaka, Tokyo 181-8588, Japan}
\altaffiltext{2}{Department of Astronomy, Graduate School of Science, University of Tokyo, 7-3-1
Hongo, Bunkyo-ku, Tokyo 113-0033, Japan}
\altaffiltext{3}{Center for Astrophysics, 
Department of Physics, 
University of Notre Dame, Notre Dame,
IN 46556, U.S.A.}
\altaffiltext{4}{Department of Physics, Tokyo Institute of Technology,
2-12-1 O-Okayama Meguro-ku, Tokyo 152-0033, Japan}
\altaffiltext{5}{Department of Astronomy and Astrophysics, LASR103, University of
Chicago, 5640 South Ellis Avenue, Chicago, IL 60637}
\altaffiltext{6}{present address : Naka Division, Nanotechnology 
Products Business Group, Hitachi High-Technologies Corporation, 882,
Ichige, Hitachinaka, Ibaraki, 312-0033, Japan}

\section{Introduction}
It is generally believed that the r-process is associated with
supernova (SN) 
explosions (Meyer et al. 1992; Woosley et al. 1994; Witti, Janka \&
Takahashi 1994), neutron star mergers (Freiburghaus, Rosswog \&
Thielemann 1999) 
or gamma-ray bursts (GRB) (Cameron 2001, 2003; Inoue et al. 2003)
in which a neutron-rich environment is realized.
However, 
the astronomical site for the r-process has not been unambiguously
determined.
Clarifying the origin of the r-process and understanding the heavy element
production therein are currently important subjects in astrophysics.

In the present work, we attempt to clarify the effects of 
nuclear physics uncertainties
in r-process nucleosynthesis  
by investigating the dependence of the r-process yields on the nuclear 
reaction rates and also on the explosion
environment. In particular, we first 
identify the most important light-mass
nuclear reactions for the r-process by using the concept of
reaction ``sensitivity''  as defined in Section \ref{sensitivity}.
We discuss the
nucleosynthesis network in Section \ref{network}, and 
describe three representative dynamic flow models 
in Section \ref{wind}.
The results of deduced reaction sensitivities are given in 
Section \ref{results}, 
where we also investigate the 
sensitivity of the most important light-element reactions  
on the explosion conditions. 
Our defined ``sensitivity'' for the efficiency of the nuclear
reactions is shown to be a useful measure of the production of actinides. 
In this context it is also relevant to cosmochronology. 
In Section \ref{carbon} an illustration is given based upon
a  detailed analysis of the waiting point for 
neutron-rich carbon isotopes.  This study highlights
both the uncertainty and importance of obtaining
experimental data, as well as the role of $(\alpha$,n) reactions
along the r-process path for light nuclei. 

\subsection{Background}

The most probable astronomical site for the r-process is in 
core-collapse supernovae (e.g. Type II, Ib, Ic SNe). This is
supported by the observational fact that the abundance pattern of 
metal deficient stars has an apparent universality, and 
is very similar to the solar abundance distribution 
(Sneden et al. 1996, Honda et al. 2004, etc.).
Metal-poor stars ([Fe/H] $\leq$ -3) 
have only been influenced by one or two supernovae.
The fact that the r-process
abundance pattern of these early objects 
is quite similar to the Solar abundance distribution
suggests that 
supernova explosions have supplied
r-process elements in a similar manner both in the past and in more recent times.

Although supernova explosions are not yet 
fully understood,  it has been established (Meyer et al. 1992; Woosley et
al. 1994; Witti, Janka \& Takahashi 1994) that 
some conditions of supernova hydrodynamics
(i.e. the hot-bubble region formed by a $\nu$-driven wind)
can produce r-process conditions.
One such condition is an environment with excess neutrons,
i.e.~the electron fraction $Y_{e}$ is less than 0.5.
Sufficient neutrons per seed nucleus are a necessary condition 
for a successful r-process. The r-process occurs effectively when the 
neutron-to-seed ratio is large.  Here, ``{\it seed}'' means nuclei
whose mass number is typically between 70 and 120 (including for
example  
$^{78}\mathrm{Ni}$).  Such seed nuclei are made through the $\alpha$-process
(Woosley \& Hoffman 1992).

In the wind models the r-process conditions are
achieved even for modest $Y_{e}$ in an
environment with high entropy. 
When the entropy per baryon is large, $s/k_{B}\geq$ 
200 (Woosley et al. 1994, Takahashi and
Janka 1997), 
the r-process can occur from the initial condition in which 
most nuclei are dissociated into alpha particles plus free neutrons and
protons.
Another condition is that of a short dynamical timescale (Otsuki et
al. 2000), where the dynamical time, $\tau_{dyn}$, 
is defined as the e-fold decay time of the temperature
from an initial value of $T_{9}=0.5$ MeV (Qian \& Woosley 1996).
A short dynamical time can suppress the overproduction of seed nuclei, 
leaving plenty of free neutrons for the subsequent neutron-capture flow.  This is 
because the temperature drops so rapidly
 that  the $\alpha$-process
becomes ineffective to produce seed.
If the dynamical timescale of the neutrino-energized
wind is as short as $\sim$ 5 ms (Otsuki et al. 2000), 
the 3rd extended r-process peak and even actinide nuclei can be produced.  

Moreover, there are now two views about the r-process in supernovae. 
In one view 
supernova explosions
are the universal site of the r-process (Qian \& Woosley 1996). 
In the other view, not only supernovae but also
some other phenomenon has contributed to the 
r-process (Qian \& Wasserburg 2000). 
Such a site might be the GRB environment (Cameron 2001, 2003; Inoue et
al. 2003), hypernovae or collapsars (Pruet, Woosley \& Hoffman 2003;
Fujimoto et al. 2004), or neutron star
mergers 
(
Freiburghaus, Rosswog \& Thielemann 1999). 
In the present work, we will take the former position, $i.e.$ that heavy
elements, including even the 3rd peak and actinide nuclei, can be
sufficiently synthesized in supernovae. In particular, we consider the
hot-bubble scenario containing $\nu$-driven winds in gravitational
core-collapse Type II supernovae.

Obviously, there remain many questions surrounding the r-process. 
In order to ultimately identify the astrophysical site for the r-process
it is essential to clarify the effects of the 
uncertainties in the input nuclear physics on the production of r-process
nuclei.  That is the aim of the present study.

\section{Sensitivity Parameter $\alpha_{i}$} 
\label{sensitivity}

\begin{figure}
\centering
\rotatebox{0}{\includegraphics[width=13.5cm,height=9.5cm]{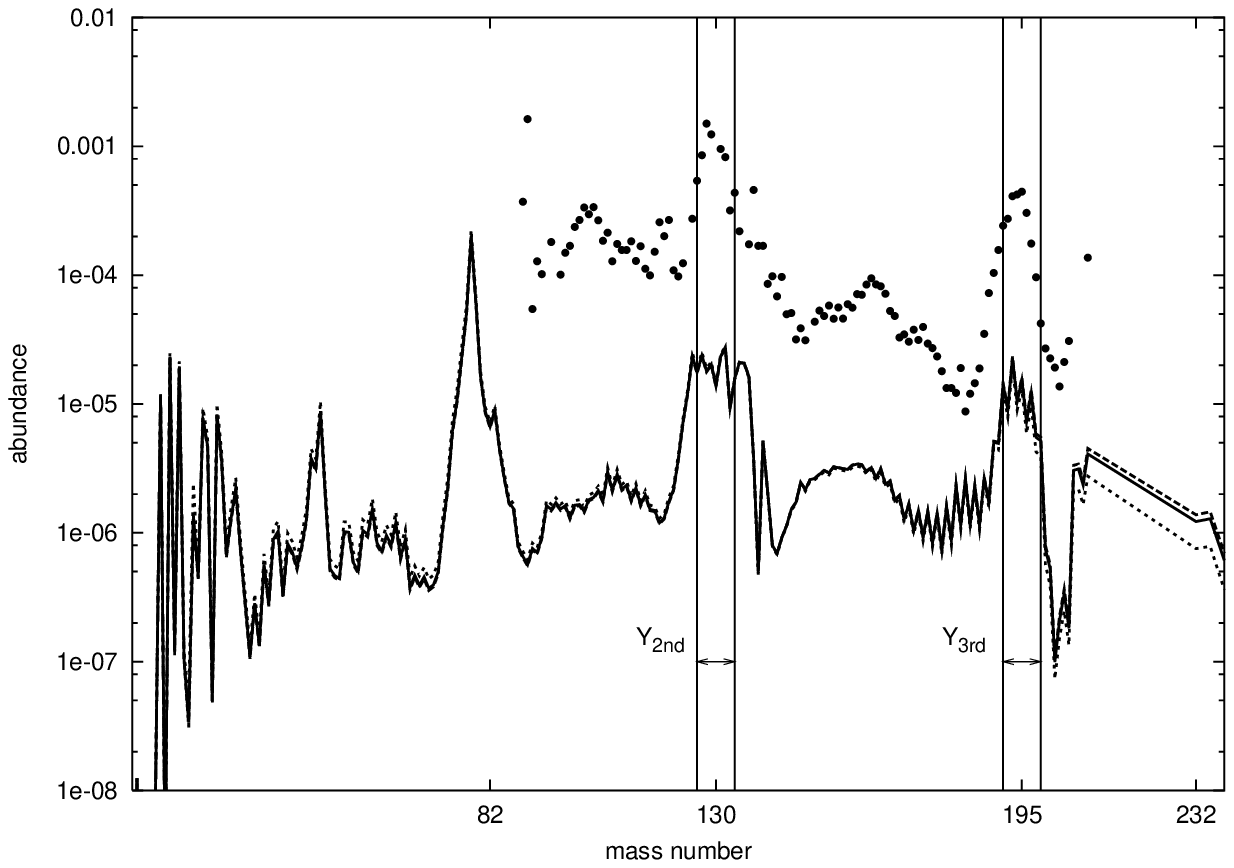}}
\rotatebox{0}{\includegraphics[width=13.5cm,height=9.5cm]{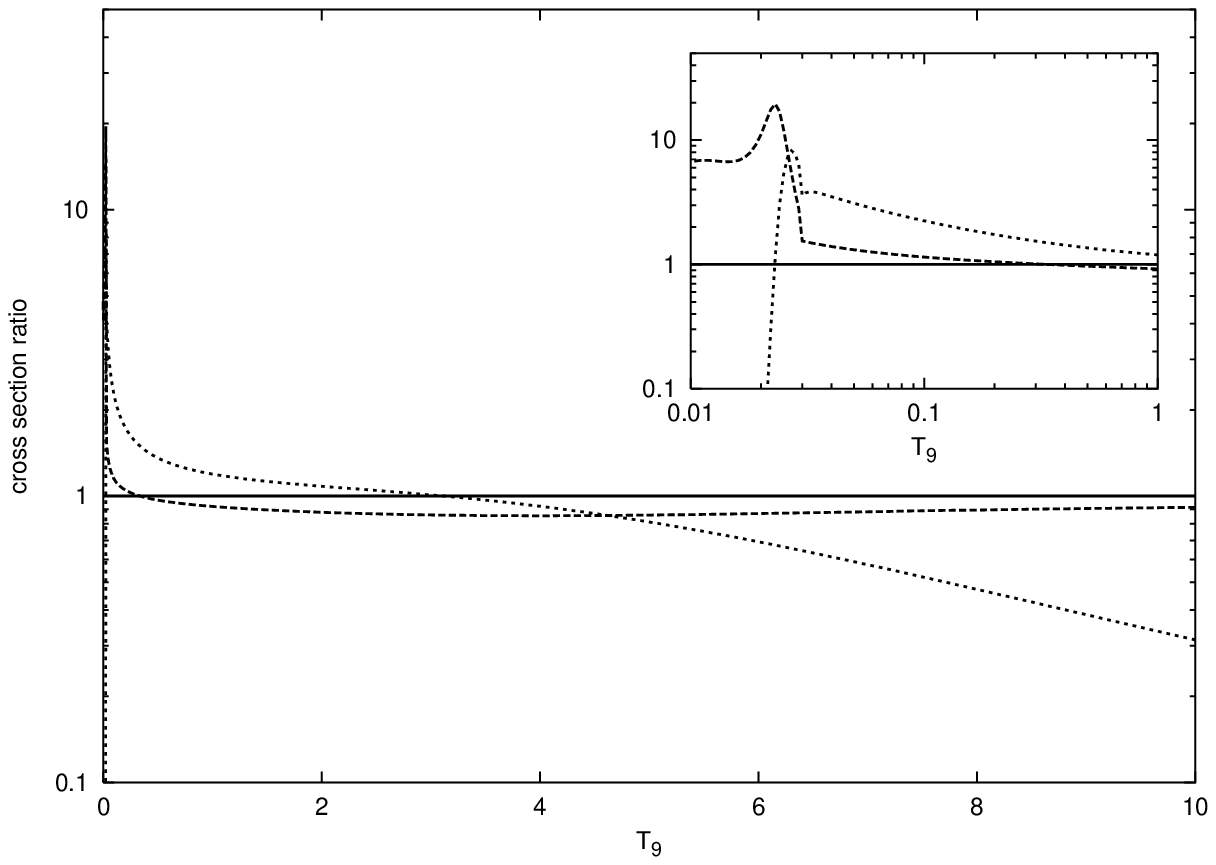}}
\caption{The upper panel shows a 
comparison of the final r-process abundances obtained with different
 $\alpha$($\alpha$n,$\gamma$)$^{9}$Be reaction rates. 
 Dotted line is for the rate of Woosley and
 Hoffman 1992 (which agrees with that of Caughlan and Fowler (1988)
 within 30 $\%$), the dashed line is for the rate of the NACRE
 compilation (Angulo et al. 1999), and 
 the solid line is for the rate of Utsunomiya et
 al. (2001) and Sumiyoshi et al. (2002). $Y_{2nd}$ and $Y_{3rd}$
 indicate  the typical r-process 
 abundances of the 2nd peak and 3rd peak elements, respectively.  The
 solar system r-process abundances from Arlandini et al. (1999) are shown
 by filled circles. 
 The lower panel shows the reaction rates as a function of temperature,
 $T_9$, as explained above. Inset highlights the lower temperatures which are
 most relevant to the r-process nucleosynthesis.}
\label{f0}
\end{figure}

In analogy with the power-law technique applied by Bahcall (1982) to
analyze the sensitivity of nuclear reactions to the Solar
neutrino flux, 
we express the calculated abundances $Y_{r}$ of r-process nucleosynthesis
 as
\begin{equation}
{{Y_{r}}\over{Y_{r}(0)}} = \prod_{i}
\left({{\lambda_{i}}\over{\lambda_{i}(0)}}\right)^{\alpha_{i}(\{\lambda_j\})},
\label{ys}
\end{equation}
where the subscript $r$ stands for typical r-process elements at the
2nd or 
3rd peak, or the actinides.  Specifically, we define 
\begin{equation}
Y_{2\mathrm{nd}} = \sum_{126 \leq \rm{A} \leq 134} Y(\rm{A}), 
\label{y2nd}
\end{equation}
\begin{equation}
Y_{3\mathrm{rd}} = \sum_{191 \leq \rm{A} \leq 199} Y(\rm{A}), 
\label{y3rd}
\end{equation}
\begin{equation}
Y_{\mathrm{actinide}} = Y(232),~Y(235),~\mathrm{or}~Y(238).
\label{yacti}
\end{equation}
As usual, $Y(A) = {n(A)/{\rho~N_{AV}}}$ 
denotes the number fraction of
element A, where $n(A)$ is the number density,
$\rho$ is the baryon mass density, and $N_{AV}$ is Avogadro's number.
$Y_{2\mathrm{nd}}$, $Y_{3\mathrm{rd}}$, and $Y_{\mathrm{actinide}}$
include the abundances of several typical r-process elements, e.g.
$^{129,131}$Xe and $^{130}$Te for the 2nd peak,
$^{194,195,196}$Pt for the 3rd peak, and $^{232}$Th, $^{235}$U, or
$^{238}$U for 
the actinides, as shown in Figure \ref{f0}. 
Although the r-process abundance peaks change slightly,
depending on the adopted hydrodynamics flow models, our conclusions 
are rather insensitive to the specific peak elements 
used in  Eqs. (\ref{y2nd})-
(\ref{yacti}). 
The $\lambda_{i}$ in Eq. (\ref{ys}) are the thermonuclear reaction rates,
\begin{eqnarray}
 \lambda_{i}&\equiv &N_{AV}~\langle \sigma_{i}^{(r)}(E)~v \rangle
\nonumber \\
&=& N_{AV}~\left(\frac{8}{\pi \mu _{12} (kT)^3}\right)^{1/2}
\int dE \sigma_{i}^{(r)}(E) E \exp \left(\frac{-E}{kT}\right),
\label{rate}
\end{eqnarray}
where $\mu _{12}$ is the reduced mass of the incident particles, $k$ is Boltzmann's
constant, and $T$ is the temperature of the gas under consideration.
$\lambda_{i}(0)$ is the adopted "standard" value of the i-th reaction rate in the reaction network as described in Section \ref{network}. Thus $Y_r(0)$ ($r$ = 2nd or 3rd) is the
r-process yield for these $\lambda_i(0)$ values. Note, that a variation
of the thermonuclear reaction rate from $\lambda_{i}(0)$ to $\lambda_{i}$ must
include the corresponding change of the reverse rate. 

The reaction cross sections $\sigma_{i}^{(r)}(E)$ have both resonant and continuum background contributions
in general. This can lead to different temperature-dependences of the thermonuclear
reaction rates. Nevertheless, most of the reactions of interest here
have only a modest temperature dependence. Hence, 
we here only take into account changes in the strength of $\lambda_{i}$, assuming the same $T$-dependence.
The power index $\alpha_i$ in Eq. (\ref{ys}) is hereafter
called the "sensitivity" parameter. 
 
\subsection{Dependence on the Astrophysical Environment}

In general, the sensitivity parameters $\alpha_i$ will depend upon the 
physical  conditions of the r-process site. 
Previous studies (e.g.~Meyer et al. 1992; Woosley et al. 1994; Qian and Woosley
1996; Otsuki et al. 2000; Terasawa et al. 2002) have identified four physical parameters which characterize the physical
conditions of the
r-process. 
%
%
These are: 1) the entropy per baryon $s/k$; 
2)  the expansion time scale $\tau_{dyn}$;
3) the initial electron fraction $Y_{e,i}$, and the asymptotic
boundary temperature $T_{a}$ of the adopted wind model.
Values of these parameters appropriate to the
models considered here will be given in the Section \ref{wind}.

Our adopted sensitivity formula, Eq.~(\ref{ys}),  has several advantages:
First, although r-process nucleosynthesis in SN explosions is 
complicated, 
the dependence of the final yields on each nuclear reaction rate is expressed in a simple separable form.
This makes the analysis of the reaction sensitivity very simple
no matter how complex the explosion dynamics are.
Secondly, Eq.~(\ref{ys}) represents even the highly
non-linear effects of the complicated reaction processes in large-scale nuclear reaction networks.
The larger the $\vert \alpha_i \vert$, the more strongly the r-process
yields depend 
on the i-th nuclear reaction rate.  If the $\alpha_i$ are close to
unity, the r-process yields are linearly dependent on that thermonuclear
reaction rate. If the $\alpha_{i}$ are close to 0, the r-process yields
are not affected. 
Note, however, that $\alpha_i = 0$ does not mean that the r-process is
totally independent of the i-th nuclear reaction. For example, 
$\alpha_i$ $\approx$ 0 can occur whenever the reaction collision time scale,
 $t_{coll} = \left( n(A) \sigma v \right)^{-1}$, is
smaller than the expansion time scale, i.e.~$t_{coll} \ll \tau_{dyn}$, 
as is the case for 
a classical r-process.  

\subsection{Incoherent Approximation}
Since $\alpha_{i}$ is a function of the $\lambda_{j}$'s, the logarithm of
Eq.~(\ref{ys})
can be written as a Taylor series expansion in powers of 
$\log\left({\lambda_{i}}/{\lambda_{i}(0)}\right)$
\begin{eqnarray}
\log\left({{Y_r}\over{Y_r(0)}}\right) &=&
\sum_{i} \alpha_{i}(\{\lambda_{j}\}) \log\left({{\lambda_{i}}\over{\lambda_{i}(0)}}\right)
\nonumber \\
&=&\sum_{i} \alpha_{i}(0) \log\left({{\lambda_{i}}\over{\lambda_{i}(0)}}\right)
\nonumber \\
&&
+ {1\over{2}}\sum_{i,j} \left({{\partial \alpha_j(0)}\over{\partial \log (\lambda_{i}/\lambda_{i}(0))}} +
{{\partial \alpha_i(0)}\over{\partial \log (\lambda_{j}/\lambda_{j}(0))}} \right)
\log\left({{\lambda_{i}}\over{\lambda_{i}(0)}}\right)
\log\left({{\lambda_{j}}\over{\lambda_{j}(0)}}\right)
\nonumber \\
&&
+ \rm{\ higher\  order\  terms},
\label{approx}
\end{eqnarray}
where $\alpha_{i}(0) \equiv \alpha_i(\{\lambda_j=\lambda_j(0)$ for all j$\})$.
The factor \\
\(\left({\partial \alpha_j(0)}/{\partial \log (\lambda_{i}/\lambda_{i}(0))} +
{\partial \alpha_i(0)}/{\partial \log (\lambda_{j}/\lambda_{j}(0))} \right)\) in the second
term does not vanish in general.
In practice, however, this coherent term between $\lambda_{i}$ and $\lambda_{j}$ (\(i
\ne j\)) is small for the adopted standard $\lambda_{i}(0)$ values.
Hence we can set 
\({\lambda_{i}}/{\lambda_{i}(0)} \approx 1\) and 
\(\log\left({\lambda_{i}}/{\lambda_{i}(0)}\right) \ll 1\).
Equation (\ref{approx}) is thus well approximated by only the first sum of the power series
\begin{eqnarray}
\log({{Y_{r}}\over{Y_{r}(0)}}) \approx \sum_{i} \alpha_{i}(0) \log\left({{\lambda_{i}}\over{\lambda_{i}(0)}}\right).
\label{approx2}
\end{eqnarray}
The validity of this incoherent approximation will be justified in 
numerical analysis below in Appendix A.

Although several important reaction cross sections of the light-mass 
nuclei have been studied both experimentally and theoretically, they 
still have appreciable uncertainties which may affect the final
r-process yield. These will be discussed in Section 3. 

The adopted range of $\lambda_{i}/\lambda_{i}(0)$ may be beyond the range to
be justified in the incoherent 
approximation (\ref{approx}). 
Nevertheless, our calculated $\alpha_i$ values for a wide range of
${\lambda_{i}/{\lambda_{i}(0)}}$ are still useful as a means to
characterize the associated changes
in $Y_{r}$ to be expected as more precise experiments or theoretical 
determinations of reaction rates become available. 

\section{Nucleosynthesis Network}
\label{network}
For the calculations of r-process nucleosynthesis, we employ the reaction
network used in Otsuki et al.~(2003), which was developed from the
original dynamical network code calculations described in Meyer et
al. (1992), Woosley et al. (1994), and Terasawa et al. (2001) so that it 
includes light neutron-rich nuclei (with Z$\leq$10). These light
neutron-rich nuclei were shown to 
be important for a successful r-process in models with a short dynamical
timescale (Terasawa et al.~2001). They play the significant roles in
the production of seed elements 
in the earlier stage of $\alpha$-rich freeze-out prior to the r-process
(Woosley \& Hoffman 1992; Meyer et al. 1992; Woosley et al. 1994). Even
so, most reaction rates of extremely neutron-rich radioactive nuclei 
are still unmeasured or poorly known experimentally. Several other
nuclear reaction cross sections at low energies 
of astrophysical interest have been measured experimentally, 
however, those are still limited to the reactions of stable and long lived
radioactive 
nuclei (e.g. Angulo et al.1999). The individual rates of the most important 
18 light-element reactions and
their adopted uncertainties are listed in Table
\ref{reaction}.
Several important modifications have been made in the present
reaction network.

\subsection{\aan}
The \aan~ reaction is of particular importance in the neutron-rich
r-process  models as has been noted by a number of authors
(Delano \& Cameron 1971; 
Meyer et al.~1992; Witti et al. 1994).  As such, 
it has been intensively studied in recent years (Angulo
et al.~1999; Utsunomiya et al.~2001; Buchmann et
al. 2001; Descouvemont 2002; Mukha et al.~2005).
 In the lower part
of Figure
\ref{f0} we show three currently used thermonuclear reaction rates for this reaction (Caughlan \& Fowler 1988 (CF88); Angulo et al.~1999
 (NACRE); Sumiyoshi et al.~2002).  In the upper part of
this figure we show the calculated r-process abundance yields,  based
upon these rates, using a wind 
model given in Otsuki et al. (2003).  Although three reaction rates are
in reasonable agreement (within factor of three) at relevant temperatures
$(0.1 \le T_{9} \le 10)$ for the r-process, the final abundance yields of
the actinide elements are very sensitive to their difference even at the
30$\%$ level at \(1 \le T_{9} \le 5\) and show appreciable differences.  
This affects 
cosmochronology based upon the ratio of long-lived radioactive elements, 
thorium (Th) and
uranium (U), to typical r-process elements such as europium
(Eu). In this specific model, for example, the
estimated age difference in
the Th/Eu-cosmochronometer (Cowan et al.~1997; Schatz et al.~2002; Otsuki
et al.~2003) turns out to be $\Delta T$ = 9.35 Gy when we apply the calculated
abundances for the CF88 and Woosley \& Hoffman (1992) rates, 
while it is only 
$\Delta T$ = 3.00 Gy for the NACRE and Sumiyoshi et al. (2002) rates.

In the present paper we take the three-body reaction rate for 
$\alpha(\alpha \rm{n},\gamma)^9 \rm{Be}$ from the network
estimate (Sumiyoshi et al. 2002) based on the experimental data
of Utsunomiya et al. (2001), which spans
the low energy region of astrophysical interest. Since direct
measurement of the radiative three-body capture reaction \aan~ is
impossible, Utsunomiya et al. (2001) measured photodisintegration cross
section of $^{9}$Be with quasi-monochromatic $\gamma$ rays produced by
means of inverse Compton scattering of laser photons. Applying the
principle of the detailed balance to the measured cross section, they
obtained the forward reaction cross section. 

Although the data include 
all relevant resonance states of astrophysical interest, there is still
$\pm$35$\%$(1$\sigma$) uncertainty in the thermonuclear reaction rates
among those of Utsunomiya et al. (2001), Angulo et al. (1999), Mukha et
al. (2005), and Buchmann et al. (2001). This arises from different
estimates of resonance parameters from ($\gamma$,n), (e,e$^{'}$), and
$\beta$-decay experiments as discussed in details by Sumiyoshi et
al. (2002).  

\subsection{\atg}
We take the reaction rate for $\alpha (t,\gamma)^{7}
\rm{Li}$ from Kajino (1986) and Kajino, Toki, and Austin (1987) as
recommended by Angulo et al. (1999). This reaction has been studied very
well by several experimental groups (Brune et al. 1994, and references 
therein) for the importance in the Big-Bang nucleosynthesis along with the
mirror conjugate $\alpha$($^3$He,$\gamma$)$^7$Be reaction for solving
the missing solar neutrino problem (Adelberger et al. 1998). The NACRE
compilation, which shows $\pm$10$\%$(1$\sigma$) uncertainty of the
$\alpha$(t,$\gamma$)$^7$Li rate, has recently been reevaluated
theoretically by Descouvemont et al. (2004) (hereafter ADNDT04) using 
the R-matrix analysis. 

Although the new ADNDT04 analysis shows even smaller uncertainty of
$\pm$4$\%$(1$\sigma$) than the NACRE compilation, more general treatment
of the error bar estimate (Smith et al. 1993) indicates
$\pm$20$\%$(1$\sigma$) uncertainty. 
A recent Monte Carlo analysis (Nollett $\&$ Burles 2000) has indicated
even larger uncertainty of $\pm$30$\%$(1$\sigma$) at lower temperature
$T_{9}$ = 0.1 - 0.2. In view of this, we adopt $\pm$30$\%$(1$\sigma$)
uncertainty in the present paper. 

\subsection{$^7$Li(n,$\gamma$)$^8$Li}
We take the reaction rate for $^7$Li(n,$\gamma$)$^8$Li from Nagai et
al. (1991b). Their measured cross section at the neutron energy of
30keV, \(\sigma (30keV) = 39.3 \pm 6.0 \mu \rm{b}\), is consistent with
the thermal reaction cross section tabulated by Mughabghab et
al. (1981), \(\sigma_{th} = 0.045 \pm 0.003 \rm{b}\), when one expects
$1/v$ extrapolation from the thermal neutron energy to 30keV. However,
Nagai's measurement is systematically larger by 30$\%$ - 50$\%$ than
those measured by Wiescher et al. (1989) in the energy range
25-420keV. We therefore estimate that the uncertainty of
$\pm$35$\%$(1$\sigma$) still remains for the $^7$Li(n,$\gamma$)$^8$Li
reaction rate at astrophysical energies. 

\subsection{$^8$Li($\alpha$,n)$^{11}$B}
The reaction rate for $^8$Li($\alpha$,n)$^{11}$B is taken from Gu et
al. (1995).  
This reaction was identified to be a critical nuclear reaction of producing
the intermediate-to-heavy mass elements in some inhomogeneous Big-Bang
nucleosynthesis models (Malaney \& Fowler 1988,\  Boyd \& Kajino 1989,\ 
Kajino \& Boyd 1990) as well as for r-process nucleosynthesis (Terasawa
et al. 2001,\ Kajino, Wanajo \& Mathews 2002). Ishiyama et al. (2004) have
recently carried out very precise measurements of the exclusive reaction cross section
for $^{8}\rm{Li}(\alpha, \rm{n})^{11} \rm{B^{\ast}}$ as well as for 
$^{8}\rm{Li}(\alpha, \rm{n})^{11} \rm{B_{gs}}$. Their result confirms 
that the transitions leading to several excited states of $^{11}\rm{B}$
make a predominant contribution to the total reaction cross section, 
which is in good agreement with the previous measurements of the inclusive
reaction cross section (Boyd et al. 1992;\ Gu et al. 1995;\ Mizoi et
al. 2000;\ Paradellis et al. 1990). 

Although each experiment reported
$\pm$20$\%$ (Gu et al. 1995), $\pm$60$\%$ (Mizoi et al. 2000), or
$\pm$30$\%$ (Ishiyama et al. 2004) uncertainty of the measured cross
section at \(0.6 \rm{MeV} \leq E\), their absolute values are very
different from one another by factor 1.7 - 2.1. We are therefore 
forced  
to estimate a factor of two uncertainty for the $^8$Li($\alpha$,n)$^{11}$B
reaction rate at astrophysically relevant energies \(E \leq 0.5 
\rm{MeV}\). 

\subsection{$^9$Be(n,$\gamma$)$^{10}$Be and $^{A}$B(n,$\gamma$)$^{A+1}$B
  (A=11-14)}
The reaction rate for $^9$Be(n,$\gamma$)$^{10}$Be,
$^{11}$B(n,$\gamma$)$^{12}$B, $^{12}$B(n,$\gamma$)$^{13}$B, and
$^{13}$B(n,$\gamma$)$^{14}$B are taken from Rauscher et
al.(1994). Although these rates were calculated by using the best available
resonance parameters,
there are still many uncertainties which arise from assumed neutron
spectroscopic factors, $\gamma$-widths, etc. It therefore is a
fair estimate that these reaction rates have at least a factor of two
uncertainty. 

There is no datum available for the $^{14}$B(n,$\gamma$)$^{15}$B reaction
rate. It is expected, however, that the radiative neutron-capture from
scattering s- and p-waves occurs in a similar manner to the
$^{13}$B(n,$\gamma$)$^{14}$B reaction at low energies of astrophysical
interest because of their similar spin-angular momentum coupling
scheme. We therefore estimate the reaction rate for
$^{14}B$(n,$\gamma$)$^{15}$B from the $^{13}$B(n,$\gamma$)$^{14}$B rate given by
Rauscher et al. (1994), with the correction due to different reaction Q
values being taken into consideration. 

\subsection{$^{A}$C(n,$\gamma$)$^{A+1}$C (A=12-19) and $^{18}$C($\alpha$,n)$^{21}$O}
The reaction rate for $^{12}$C(n,$\gamma$)$^{13}$C is taken from Nagai et
al. (1991a). Gibbons et al. (1961) old uncertain datum, \(\sigma (30keV)
= 200 \pm 400 \mu \rm{b}\), was revised by precise experimental values
of Nagai et al. (1991a), \(\langle \sigma \rangle = 16.8 \pm 2.1 \mu
\rm{b}\), and Ohsaki et al. (1994), \(\langle \sigma \rangle = 15.4 \pm
1.0 \mu \rm{b}\), which are in reasonable agreement with each other
within their 1$\sigma$ error bars, where $\langle \sigma \rangle$ is the
Maxwellian-averaged neutron-capture cross section at the temperature
\(kT = 30keV\). They also decomposed direct p-wave and d-wave
contributions to the non-resonant neutron capture cross sections
(Kikuchi et al. 1998). These new data are also consistent with the
measured upper limit of Macklin (1990), \(\langle \sigma \rangle < 14
\mu \rm{b}\). Careful analysis of the combined precise data of Nagai et
al. (1991a) and Ohsaki et al. (1994), so as to minimize the
reduced-$\chi ^2$ by taking account of its error from the $\chi ^{2}$
distribution, indicates $\pm$10$\%$(1$\sigma$) uncertainty of the
$^{12}$C(n,$\gamma$)$^{13}$C reaction rate.

Raman et al. (1990) measured the $\gamma$-width of the 2$^{+}$ state of
$^{14}$C which lies at 142 keV above the $^{13}$C + n particle
threshold. Rauscher et al. (1994) assumed predominance of this
resonance contribution to the 
$^{13}$C(n,$\gamma$)$^{14}$C cross section. Since the measured $\gamma$-width has
+40$\%$ and -16$\%$ error bars (Raman et al. 1990) and the other
resonances or direct capture component may contribute to the total
reaction rate, we estimate a factor of two uncertainty of the reaction
rate for $^{13}$C(n,$\gamma$)$^{14}$C.  

The $^{14}$C(n,$\gamma$)$^{15}$C cross section at astrophysical low
energies was measured by Beer et al. (1992) in direct neutron-capture
experiments, and also by Howath et al. (2002) and Nakamura et al. (2003) via
Coulomb dissociation of $^{15}$C.  These authors report respectively
$\pm$25$\%$,  $\pm$35$\%$, and $\pm$10$\%$ uncertainties of their
Maxwellian-averaged neutron-capture cross sections at the temperature
\(kT = 23keV\). For its importance in inhomogeneous Big-Bang
nucleosynthesis of intermediate-to-heavy mass nuclei (Wiescher, Gorres\&
Thielemann 1990; Kajino, Mathews \& Fuller 1990), another Coulomb
dissociation experiment (Pramanik et al. 2003) was carried out. The
result is
more consistent with the direct neutron-capture
experiment. Absolute values of these four reaction rates unfortunately
differ by a factor of 2-4 from one another. Therefore we estimate a factor
of four uncertainty of the reaction rate for
$^{14}$C(n,$\gamma$)$^{15}$C. 

The reaction rates for $^{15}$C(n,$\gamma$)$^{16}$C and
$^{16}$C(n,$\gamma$)$^{17}$C are taken from Rauscher et
al. (1994). These rates have at least a factor of two uncertainty for the
same reasons discussed above for the $^9$Be(n,$\gamma$)$^{10}$Be and
$^{A}$B(n,$\gamma$)$^{A+1}$B (A = 11, 12, and 13) reactions.

For the other neutron-rich carbon isotopes 
we make use of
Hauser-Feshbach (HF) estimates for the (n,$\gamma$) and ($\alpha$,n)
rates, as has been done in previous
studies (e.g. Terasawa et al. 2001).
We utilize these estimates even though it is generally believed that the
HF model 
is a poor approximation for light nuclei (as well as for exotic neutron
rich nuclei) 
due to the low density of states for these nuclei.  Nevertheless, 
HF estimates are easy to obtain and are 
often within a factor of two (Mathews et al 1983) of the
true cross sections.  Direct capture cross sections 
would provide a lower limit
to the cross sections (Mathews et al. 1983)  if the relevant 
spectroscopic factors were known. 
However, such nuclear data for most of the neutron-rich nuclei 
of interest here have yet to 
be obtained. As such, we consider simple HF estimates 
as good as can be obtained at the present time.
They are consistent with other network applications in the field,
and provide an adequate basis on which to judge which
cross sections have the most sensitivity for r-process nucleosynthesis.
We note that only the reaction rate for $^{18}$C(n,$\gamma$)$^{19}$C is
estimated from the measured cross section using the Coulomb dissociation
method by Nakamura et al.(1999).

Extensive neutron-rich nuclei and their associated nuclear reactions
were also added (Nakamura et al. 1994; Fukuda et al. 2004; Otsuki et
al. 2003, Terasawa et al. 2001, 2002) so that the updated network
applies to many different kinds of wind models. Wanajo et al. (2002)
suggested that the low-mass neutron-rich 
nuclei would be less
important in their specific models than that found by Terasawa et
al.~(2001) for a low adopted asymptotic temperature. As we will
discuss in the next sections, however, these nuclei play 
an important role in any wind models. 
We finally note that we calculate the nucleosynthesis sequence from the
NSE, $\alpha$-process, $\alpha$-rich freeze-out, r-process and subsequent beta-decay and alpha-decay in a single network code rather
than to split the calculation into two parts as was done in Woosley et
al. (1994). These modifications are important for our studies of the
nuclear reaction sensitivity of r-process nucleosynthesis.

\section{Neutrino-Driven Wind Models}
\label{wind}

Our purpose is to clarify the dependence of r-process
nucleosynthesis on individual nuclear reaction rates. 
The sensitivity
parameters $\alpha_{i}$, however, depend on thermodynamic quantities specific to 
the explosion conditions. 
Possible astrophysical sites for  the
r-process include  Type II SNe, neutron-star mergers, 
and several others
(e.g. Mathews and Cowan 1990).  The detailed
explosion dynamics for these sites, however, are not yet quantitatively known.

Even in the presently popular paradigm of neutrino-driven winds from Type II SNe,
the physical conditions of the r-process  are dependent
on details of the adopted numerical simulation (Meyer et al. 1992;\ 
Woosley et al. 1994;  Witti et al. 1994;\ 
Takahashi et at. 1994; Qian \& Woosley 1996;  Cardall\& Fuller
1997; Hoffman et al. 1997; Otsuki et al. 2000, 2003;  Sumiyoshi 2000; 
Thompson, Burrows \& Meyer 2001; Wanajo et al. 2001). 
In this article, therefore, we adopt three different schematic parameterizations
of the r-process conditions.
One is a fast steady-state  
neutrino-driven wind (Otsuki et al.~2000); the other is a slow
hydrodynamic wind (Woosley et al.~1994; Qian \& Woosley 1996).
The third is an 
exponential model as used in Otsuki et al.~(2003).

\subsection{Fast Steady-State Wind Model}
\label{fastflow}
One model we consider is that of a fast spherical steady-state
flow model for the neutrino-driven wind (Qian \& Woosley 1996; Takahashi \& Janka 1997;  Otsuki et al. 2000, 2003;  Thompson et al. 2001; 
  Wanajo et al.~2001).
Otsuki et al. (2000) and Sumiyoshi et al. (2000) 
found that the r-process can occur in such neutrino-driven winds
even for moderate entropies ($s/k \approx$ 100-300)
as long as the dynamical expansion timescale becomes much shorter than the collision timescale of neutrino-nucleus interactions.
We note that there has been some discussion (e.g. Thompson et al. 2001)
of a very fast r-process with high $Y_e \ge 0.485$.  In such an extreme
alpha-rich environment it is possible that the $3 \alpha$ reaction competes  with the $\alpha(\alpha,n)^9$Be reaction to assemble heavy nuclei.  In this case the r-process yields might also have some sensitivity to the $3\alpha$ reaction.  However, we do not investigate that possibility here. 

For the present application, such hydrodynamic flow can be
approximated (Otsuki et al. 2003) by solving the following non-relativistic equations:
\begin{equation}
4 \pi r^2 \rho v = \dot{M}~,\\
\label{flow_mass}
\end{equation}
\begin{equation}
{1\over 2}v^2 - {{GM}\over{r}} + N_{AV}~s_{rad}~ \mathrm{kT}= E~,\\
\label{flow_ene}
\end{equation}
\begin{equation}
s_{rad} ={{11 {\pi}^2}\over{45\rho N_{AV}}}\left({{\mathrm{k}T}\over{\hbar
c}}\right)^3~,
\label{flow_s}
\end{equation}
where $\dot{M}$ is the rate at which matter is ejected by neutrino
heating from the surface of the proto-neutron star.  

The total energy $E$ in Eq.~(\ref{flow_ene}) is
fixed by a boundary condition on the asymptotic temperature $\mathrm{T}_{a}$
\begin{eqnarray}
 E=N_{AV}~s_{rad}~\mathrm{k}~\mathrm{T}_{a}~~.
\label{energy}
\end{eqnarray}
We only take into account photons and $e^{\pm}$ pairs for relativistic
particles in the estimate of the entropy, $s_{\rm{rad}}$, in
Eq. (\ref{energy}). 
For simplicity, we utilize an adiabatic,
constant-entropy wind rather than to compute the neutrino heating
explicitly (Otsuki et al. 2000,  Wanajo et al. 2001). 
We assume a constant
neutrino luminosity of $L_{\nu}=10^{51} \rm{ergs ~s^{-1}}$ for each neutrino species. 

This model has four parameters. They are the neutron-star
mass, entropy, boundary temperature, and mass-loss rate.
We here adopt one typical flow which characterizes the ``fast wind'' with a
short dynamical explosion timescale, $\tau_{dyn}=5\ \rm{ms}$, 
and moderately high entropy per baryon, $s/{\rm k}=300$.
Parameters for this model are summarized in Table \ref{flowmodel}. 
As shown in Otsuki et al. (2003), this parameter set well reproduces 
the universal abundance pattern of neutron-capture elements.

\subsection{Slow Wind Model}

A dynamical Type II SN explosion and the subsequent r-process in the
neutrino-driven wind above the proto-neutron star were calculated by
Woosley et al. (1994). Their numerical results were based upon the spherically
symmetric hydrodynamic code described in Wilson \& Mayle (1993).
That calculation included a number of refinements over previous treatments of
the hydrodynamics and convection so that the evolution could be followed
to more than 10 s after the core bounce. Such late times are necessary
to describe properly 
the mass-loss rate and neutrino excess in the wind.  Although this
numerical simulation demonstrated
for the first time that the r-process could occur in delayed
neutrino-heated SNe, several difficulties were subsequently
identified. First, independent non-relativistic numerical models 
(e.g.~Witti et al. 1994,\ Takahashi et al. 1994) 
had difficulty producing the
required entropy $s/k \sim 400$ as indicated by Woosley et al.
(1994). Second, even should the entropy be high enough, the effects of
neutrino absorption \(\nu_{e} + n \rightarrow p + e^{-}\) and \(\nu_{e} +
A(Z,N) \rightarrow A(Z+1, N-1) + e^-\)  decrease the neutron
fraction and prohibit the r-process (Meyer 1995). 
Otsuki et al. (2000) showed that this is generally the case when the
expansion of the neutrino-driven wind is slow. 

Nevertheless, this ``slow wind'' model of Woosley et al. (1994) is still
attractive,  as a source
for the 1st and 2nd r-process peaks.
We therefore adopt the flow model of trajectory 40 in Woosley et al. (1994)
as a typical ``slow wind'' model with high entropy $s/k \sim
400$ and a large dynamical expansion timescale \(\tau_{dyn} \approx 300\
\rm{ms}\). The characteristic parameters of this wind model are 
compared with those of the ``fast wind'' model in Table \ref{flowmodel}. 

\subsection{Exponential Model}
\label{exponential}

We also utilize an exponential model similar to that employed by
Meyer and Brown (1997) in order to 
roughly approximate conditions for the r-process which
might occur in lower-entropy, low $Y_e$ environments such as
prompt SNe (Hillebrandt et al. 1984, Sumiyoshi et al. 2001,
Wanajo et al. 2003) or neutron-star mergers 
(Freiburghaus, Rosswog, \& Thielemann 1999). 

Under the condition that 
\begin{equation}
s ={{11 {\pi}^2}\over{45\rho N_{AV}}}\left({{\mathrm{k}T}\over{\hbar
c}}\right)^3=\mathrm{const.}~~,
\end{equation}
the temperature and matter density are given by
\begin{eqnarray}
T_{9}(t)=T_{9}(0)\exp(-t/{\tau}_{dyn})+T_{a},\\
\rho(t)=\rho(0)(T_0/T)^{3}\\
\rho(0)=3.3\times 10^8 {\mathrm{gcm}}^{-3},
\end{eqnarray}
where $T_{9}(0)=8.40$,
and $T_a$ is the asymptotic temperature (Terasawa et al. 2002, Otsuki et al. 2003) of the wind as given in  Eq. (\ref{energy}). This
temperature 
determines the freeze-out of neutron-capture flow. Given the adiabatic
entropy per baryon $s/{\rm k}$, the matter density evolves so that it
satisfies Eq. (\ref{energy}) where one replaces $s_{rad}$ by $s$. 
In a more realistic calculation with fixed entropy there will be an increase in temperature at the end of the expansion as electron pairs annihilate (e.g. Woosley et al. 1994).
  However, given the
schematic nature of the exponential model employed here,
we ignore that 
effect on the temperature evolution.

The parameter sets we use in the exponential models
are summarized  in Tables
\ref{flowparametertau}- \ref{flowparametertemp}.
Because sensitivities can only be meaningfully compared between
models in which a similar  abundance pattern is produced,
we have selected parameter sets which lead to nearly
the same final abundance pattern, as displayed in Figure \ref{f0}, while simultaneously varying two 
parameters.
The three models with fixed $Y_{e}$ and $T_a$ in
Table \ref{flowparametertau} thus 
give the dependence of the sensitivity on a simultaneous
variation of the dynamical timescale
$\tau_{dyn}$ and entropy $s/k$. Similarly, 
Tables \ref{flowparameterentro} and \ref{flowparameterye}
with fixed $\tau_{dyn}$ and $T_a$ show effects of a simultaneous
variation of $s/k$ and $Y_e$.  Note that we fix $\tau_{dyn}$ at
a different value in Table \ref{flowparameterentro} ($\tau_{dyn}$ = 1.0~ms)
and Table  \ref{flowparameterye} ($\tau_{dyn}$ = 5.0~ms) in order to
avoid almost degenerate choice of the two parameters $s/k$ and $Y_{e}$. 
Table \ref{flowparametertemp} 
with fixed $\tau_{dyn}$ and $Y_e$ then involves a simultaneous
variation of $s/k$ and $T_a$.


It is important to consider such a wide variety of models. 
For example, neutron star mergers (Freiburghaus, Rosswog \& Thielemann 1999) have a low $Y_{\mathrm{e}}$ $\le$ 0.2,  
while $Y_{e}$ typically  
ranges from 0.42 to 0.47 in the $\nu$-driven winds 
of core-collapse supernovae (Woosley et al. 1994). 
Although such merger models
involve even lower entropy and $Y_e$
than the models considered here, these
parameter sets, nevertheless, provide some indication as to
 the dependence of the r-process sensitivities
as $Y_e$ is decreased.

\section{Results}
\label{results}
\subsection{Most Important Light Nuclear Reactions}
In network calculations, the abundance evolution can be written as 
\begin{eqnarray}
 {{{\rm d}\mathrm{Y}_{i}}\over{{\rm d}t}}=-\sum_{j\cdots
 k}\lambda_{ij\cdots k}^{lm\cdots n}
 \mathrm{Y}_{i} \mathrm{Y}_{j} \cdots\mathrm{Y}_{k}
 +\sum_{lm\cdots n}\lambda_{lm\cdots n}^{ij\cdots k}
 \mathrm{Y}_{l} \mathrm{Y}_{m} \cdots\mathrm{Y}_{n}~~,
\end{eqnarray}
where $\mathrm{Y}_{i}$ is the number abundance, and $\lambda_{ij\cdots
k}^{lm \cdots n}$ and  $\lambda_{lm\cdots n}^{ij \cdots k}$ are the
destruction ($i+j+\cdots+k \rightarrow l+m+\cdots+n$) and production
($i+j+\cdots+k \leftarrow l+m+\cdots+n$) of the i-th nucleus,
respectively.  
To identify which reactions are most important in the r-process
 we define the flux at a particular
nucleus, j, as 
\begin{eqnarray}
F_{ij}(t)=\int_{0}^{t} \left[{{{\rm d}Y_{i}}\over{{\rm d}t^{\prime}}}(i \rightarrow j)
-{{{\rm d}Y_{j}}\over{{\rm d}t^{\prime}}}(j \rightarrow i)\right] {\rm d}t^{\prime}~~.
\end{eqnarray}
The $F_{ij}(t)$ is the net nuclear flux through the nucleus j from
\(t=0\) to a time $t$. By setting \(t \ge t_{f}\), where $t_{f}$ is the
freezeout time of the r-process, we can quantitatively 
determine which
nuclear reaction experiences the largest flux throughout the nucleosynthesis. 
We define $t_{\alpha}$ and $t_{n}$ at which the temperature of the
expanding wind becomes \(T_{\alpha} = 0.5 MeV\) and \(T_{n} = T_{\alpha}
e^{-1}\), respectively. $T_{\alpha}$ is the typical temperature at the
beginning of the $\alpha$-process when the system drops out of the
nuclear statistical equilibrium (NSE), and $T_{n}$ is the temperature
near the end of the $\alpha$-process. We can define the expansion time
scale $\tau_{dyn}$ in Eq. (13) as a typical duration time for the
$\alpha$-process nucleosynthesis, i.e. \(\tau_{dyn} = t_{n} -
t_{\alpha}\). After $t_{n}$, successive neutron-capture flow on
intermediate-to-heavy mass unstable nuclei is established. 

In this way we have 
found three equally important main-flow paths  among the
light-mass 
nuclei.  These paths predominate in all of the models considered here.
They are,  
\begin{eqnarray}
\alpha(\alpha \rm{n}, \gamma)^{9}\rm{Be} 
(\alpha,\rm{n})^{12}\rm{C}(\rm{n},\gamma)^{13}\rm{C}(\rm{n},\gamma)
{^{14}\rm{C}} (\rm{n},\gamma)^{15}\rm{C},\nonumber\\
\alpha(\alpha \rm{n}, \gamma)^{9}\rm{Be} 
(\rm{n},\gamma)^{10}\rm{Be}(\alpha,\gamma)
{^{14}\rm{C}} (\rm{n},\gamma)^{15}\rm{C},\\
\alpha(\rm{t},\gamma)^{7}\rm{Li}(\rm{n},\gamma)^{8}\rm{Li}(\alpha,\rm{n})^{11}\rm{B}(\rm{n},\gamma)^{12}\rm{B}(\rm{n},\gamma)^{13}\rm{B}(\rm{n},\gamma)^{14}\rm{B}(\rm{n},\gamma)^{15}\rm{B}(e^{-}\nu)^{15}\rm{C}.
\end{eqnarray} 
This is consistent with the results of Terasawa et al. (2001). We
have also  identified the main flow-paths to nuclei beyond $^{15}$C,\\ 
\begin{eqnarray}
^{15}\rm{C}(\rm{n},\gamma)^{16}\rm{C}
(\rm{n},\gamma)^{17}\rm{C}(\rm{n},\gamma)^{18}\rm{C}
(\alpha,\rm{n})^{21}\rm{O}(\rm{n},\gamma)^{22}\rm{O}(\rm{n},\gamma)
{^{23}\rm{O}}(\rm{n},\gamma)^{24}\rm{O}\nonumber\\
^{15}\rm{C}(\rm{n},\gamma)^{16}\rm{C}
(\rm{n},\gamma)^{17}\rm{C}(\rm{n},\gamma)^{18}\rm{C}
(e^{-}\nu)^{18}\rm{N}(\rm{n},\gamma)\cdots ^{23}\rm{N}(e^{-}\nu)
{^{23}\rm{O}}(\rm{n},\gamma)^{24}\rm{O}\nonumber\\
^{15}\rm{C}(\rm{n},\gamma)^{16}\rm{C}
(e^{-}\nu)^{16}\rm{N}(\rm{n},\gamma)^{17}\rm{N}(\rm{n},\gamma) 
{^{18}\rm{N}}(\rm{n},\gamma)\cdots ^{23}\rm{N}(e^{-}\nu)
{^{23}\rm{O}}(\rm{n},\gamma)^{24}\rm{O}\nonumber\\
\longrightarrow 
{^{24}\rm{O}}(\alpha,\rm{n})^{27}\rm{Ne}\cdots.\nonumber \\
{^{24}\rm{O}}(e^{-}\nu)^{24}\rm{F}\cdots.
\end{eqnarray}
The first nuclear reaction chain through $^9$Be
and  $^{12}$C in Eq.~(17) is well
studied in the literature (Woosley \& Hoffman 1992, Woosley et al.1994; Witti et al.~1994; Otsuki et al.~2000; Terasawa et al.~2001).  
It plays an
important role in the $\alpha$-process. The second nuclear reaction
chain through
 $^{10}$Be in Eq.~(17) and the chain of Eq.~(18) are
the newly identified main flow paths along  with the
extended flow of Eq.~(19). The significance of the flows
identified in Eqs.~(18) and
(19) is universal and will be further discussed below. 

\subsection{Reaction Sensitivity in Fast and Slow Wind Models}
We carried out network calculations and obtained the final abundances of
r-process elements. 
The 18 most relevant reactions in Table 
\ref{reaction} were analyzed to derive the sensitivity parameters as
formulated in Section \ref{sensitivity}.  
These results are summarized in Tables \ref{result1} and
\ref{result2}. 
Shown in the last two columns are the current importance of each
reaction at $\pm$2$\sigma$ uncertainties in the r-process
nucleosynthesis, which is defined by 
\begin{eqnarray}
 {{\rm{Y}_{r}}\over{Y_{r}(0)}} \equiv  
 {{\rm{Y}_{r}(0) + \delta \rm{Y}_{r} }\over{\rm{Y}_{r}(0)}} \approx
 \Bigl(1 + {{\delta
 \lambda_{i}}\over{\lambda_{i}(0)}}\Bigr)^{\alpha_{i}}, 
\end{eqnarray}
where we set \(\delta \lambda_{i} = \pm2\sigma_{i}\lambda_{i}(0)\) with
$\sigma_{i}$ tabulated in Table \ref{reaction}. We here take the average
of $^{235}$U and $^{238}$U as typical of r-process yields.  
These can be used to  
infer which reactions most strongly affect the abundance
pattern of heavy r-process elements.

\subsubsection{Fast-flow models}
\begin{figure}
\centering
\rotatebox{0}{\includegraphics[width=15cm,height=18cm]{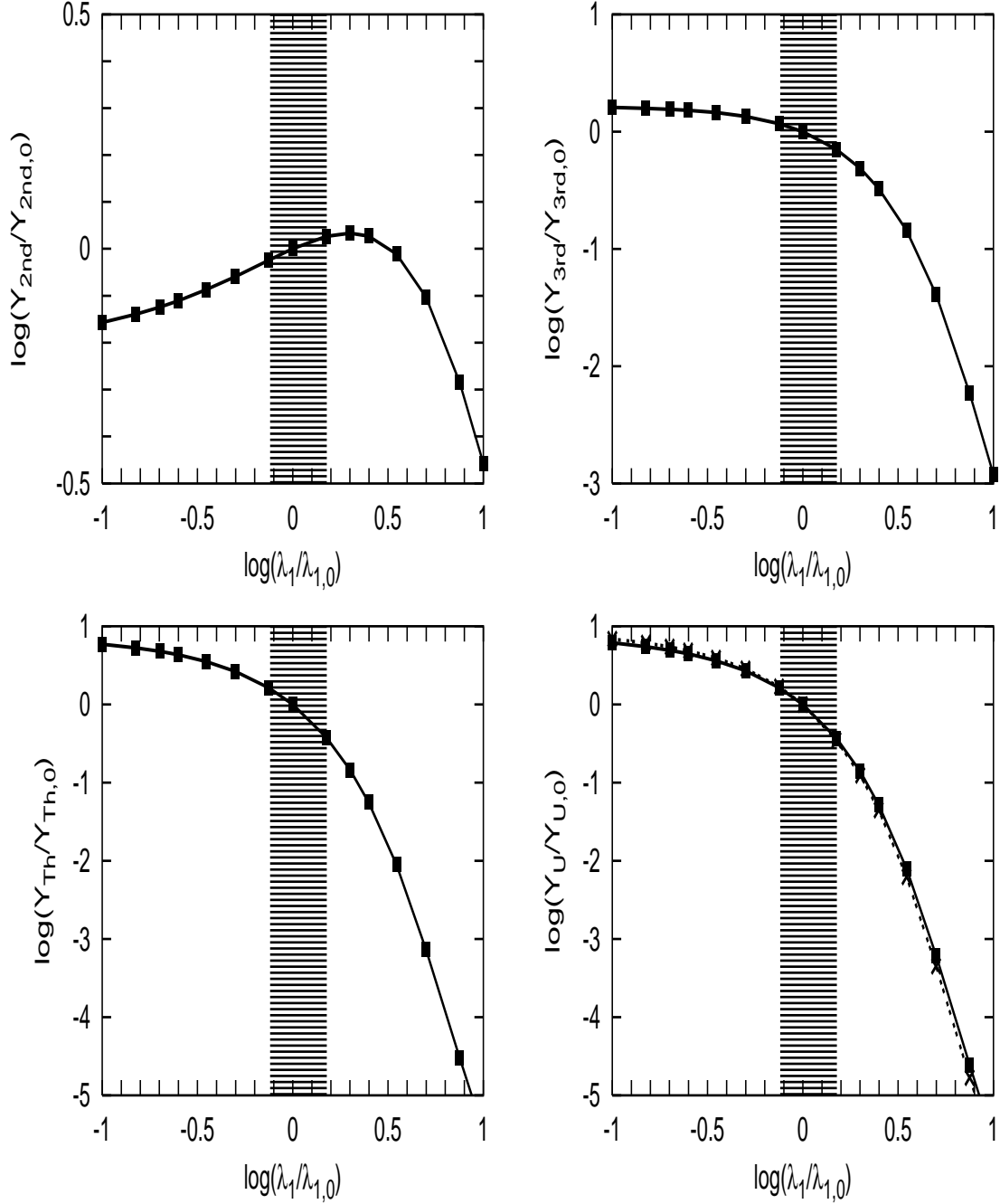}}
\caption{Abundance ratios of the 2nd-peak (upper left), 3rd-peak (upper
 right), Th (lower left), and U (lower right) elements relative to the
 original abundances  
 as a function of $\lambda_{1}/\lambda_{1}(0)$ for the
$\alpha(\alpha$n,$\gamma) ^9$Be reaction
 (1) in Table \ref{reaction}. Solid lines show the best fit to these data based on
 Eq.(7). Note that the solid and dotted lines for U (lower right)
 are for $^{235}$U and $^{238}$U, respectively.  
 The shaded areas indicate the adopted uncertainties in the nuclear 
 reaction rate tabulated in Table \ref{reaction}.}
\label{f1}
\end{figure}

We first discuss the results for the fast flow model
(Sect.~\ref{fastflow}). We show in Fig.~\ref{f1} an example of how to
determine the sensitivity parameter $\alpha_i$ 
for the $\alpha(\alpha \rm{n}, \gamma)^9$Be reaction. 
The top left, top right, lower left, and lower right panels show the
abundance ratios for 
the 2nd peak, 3rd peak, Th, and U elements, respectively, as
defined by Eqs.(1) - (4). 
Shaded regions 
indicate the $\pm$1$\sigma$ uncertainties in the reaction rate for \aan,  
as given in Table 
\ref{reaction}, and 
$\log{\left({\lambda_{i}}/{\lambda_{i}(0)}\right)}
\equiv \log\left([{\lambda_{i}(0) + \delta
\lambda_{i}}]/{\lambda_{i}(0)}\right)$. 
The sensitivity parameter $\alpha_{i}$ is the
logarithmic derivative of \(\rm{Y}_{r}/\rm{Y}_{r}(0)\) at
\(\lambda_{i}/\lambda_{i}(0) = 1\), thus the variation of
\(\log(\rm{Y}_{r}/\rm{Y}_{r}(0))\) in the shaded region is linerly
approximated by Eq.(7) very well once we calculate \(\alpha_{i} \approx
\alpha_{i}(0)\). 

The abundance for the 2nd peak (in the top left panel) has a maximum value at
$\log{\lambda_{1}/\lambda_{1}(0)}=0.35$, while the abundance for the 3rd peak 
(in the top right panel) is a monotonically decreasing function of 
$\lambda_{1}/\lambda_{1}(0)$.
Increasing $\lambda_{1}$ makes the
$\alpha$-capture process more important.
This leads to an increase in the abundance of seed nuclei,
which are mainly produced by the $\alpha$-process, and 
a decrease in the abundance of free neutrons at the end of the
$\alpha$-process. This 
decrease in the neutron-to-seed ratio is not favorable for the 
r-process. It is the reason
that the abundances of both the 2nd and 3rd peak elements decrease for $\log({\lambda_{1}}/{\lambda_{1}(0)})$ $\geq 0.35$.
On the other hand, only the abundance for the 2nd peak
decreases toward smaller $\lambda_{1}/\lambda_{1}(0)$ in the region of
$\log({\lambda_{1}/\lambda_{1}(0)}) \leq 0.35$.
Decreasing $\lambda_{1}$ means
that the $\alpha$-process becomes inefficient.
This results in a decrease
in the abundance of seed nuclei, increasing the neutron-to-seed ratio.


Reactions (2) and (4), i.e. \(\alpha(t,\gamma) ^{7} \rm{Li}\) and
\(^{8} \rm{Li}(\alpha,\rm{n}) ^{11}\rm{B}\), 
 exhibit a similar trend to that of the 
\(\alpha(\alpha \rm{n}, \gamma) ^{9} \rm{Be}\) reaction.
The reason for this is the same as for that reaction.
The neutron-capture reactions (3), (5), (7), (9), (11), (13) and (15)
show similar results to one another as expected from the calculated
sensitivity parameters in Table \ref{result1}.
As $\lambda_{i}$ for these rates  increases,
the abundances of the 2nd peak increase while
the 3rd peak decreases.
As $\lambda_{i}$ increases, these neutron-capture processes more efficient.
Neutrons are therefore exhausted at an earlier
stage and heavier elements do not form.
Thus, the abundances of the 3rd peak decrease.
On the other hand, abundances of the 2nd peak increase with $\lambda_{i}$
because nuclei capture neutrons efficiently up to
the N=82 closed shell.

The r-process has almost no sensitivity to
the n-capture reactions (8), (10), (12), (14), (16) and (17).
Therefore, variation of these reactions within our adopted 
uncertainties has almost no effect on the r-process abundances.
On these reaction paths, there are no important reactions other than
neutron-capture.


\subsubsection{Slow-flow models}
The sensitivity to charged-particle reactions
is relatively higher than the neutron-capture reactions in the slow flow
model (Sect.4.2) as indicated in Table \ref{result2}. 
As described above, the $\alpha$-process is very sensitive
to the temperature.
In the slow model,
the $\alpha$-capture process lasts for a longer time than the fast flow
model. This is because the temperature is kept high enough for
nuclei to overcome the Coulomb barrier.
Therefore, in the slow models, the 
reaction flow passes close to the $\beta$ stability line.

Comparing the results of the two flow models, we find that the sensitivities
are very dependent on the expansion timescale. 

\subsection{Sensitivity in Exponential Models}
In this subsection we discuss 
the dependence of the reaction sensitivities on a wide range
of dynamical conditions for the r-process based upon
the exponential model described in Sect.~\ref{exponential}.

\begin{figure}
\centering
\rotatebox{0}{\includegraphics[width=18cm,height=15cm]{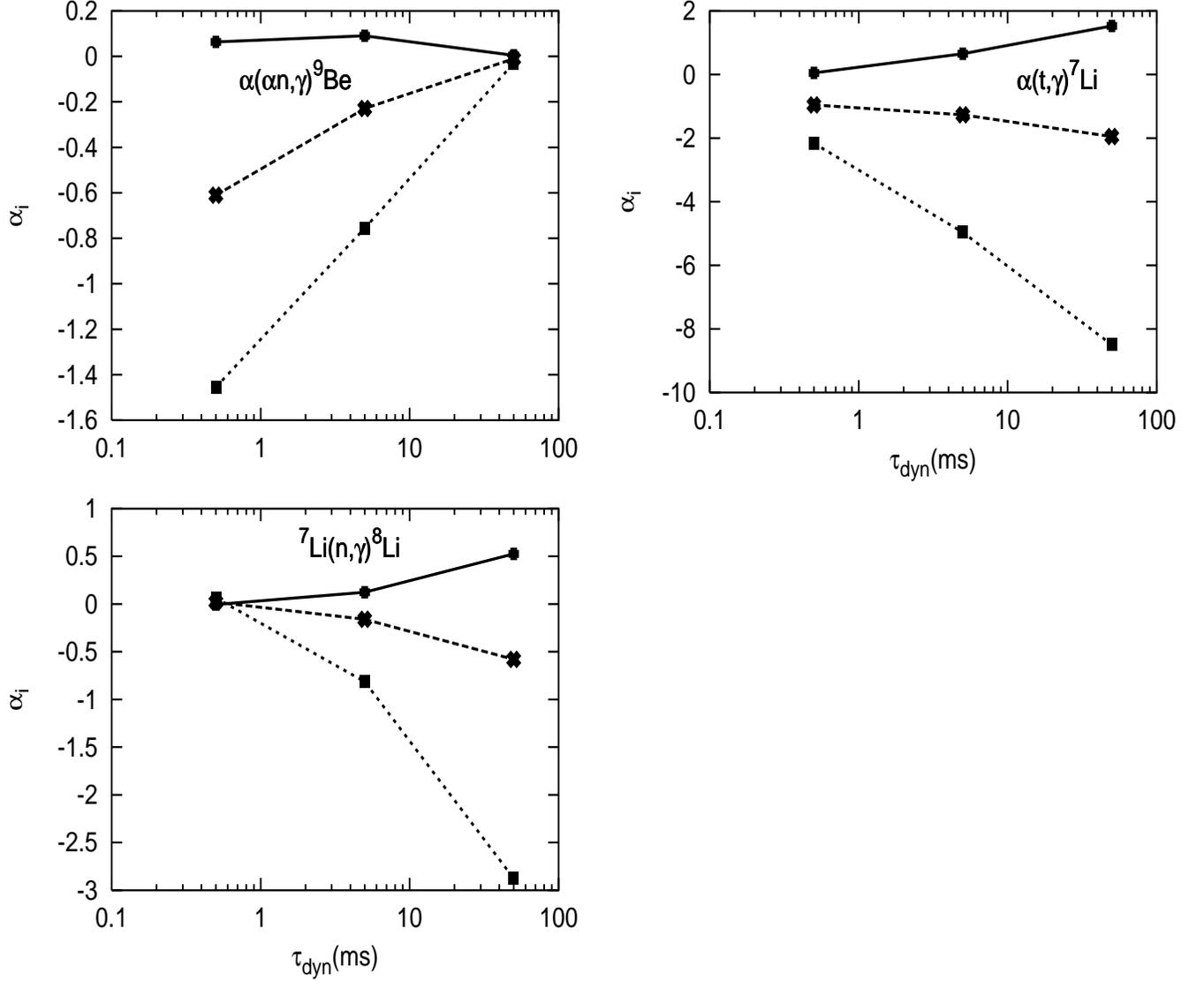}}
\caption{Dependence of the sensitivity parameter, $\alpha_{i}$, on the
 dynamical timescale,  
 $\tau_{dyn}$, for the 2nd-peak (solid line and thick-plus points), 
 3rd-peak (dashed line and thick-cross points), and
 actinide (dotted line and square points) elements, 
 corresponding to the parameter sets given in
 Table \ref{flowparametertau}. $\alpha_{i}$ for actinide elements is the
 simple average of those for $^{232}$Th, $^{235}$U, and $^{238}$U. The
 top left is for \(\alpha(\alpha 
 \rm{n}, \gamma)^{9}\rm{Be}\),   
 the top right is for \(\alpha(\rm{t},\gamma)^{7}\rm{Li}\), and the
 lower is for 
 \( ^{7}\rm{Li}(\rm{n},\gamma)^{8}\rm{Li}\).} 

\label{fgtau_nama}
\end{figure}

\begin{figure}
\centering
\rotatebox{0}{\includegraphics[width=18cm,height=15cm]{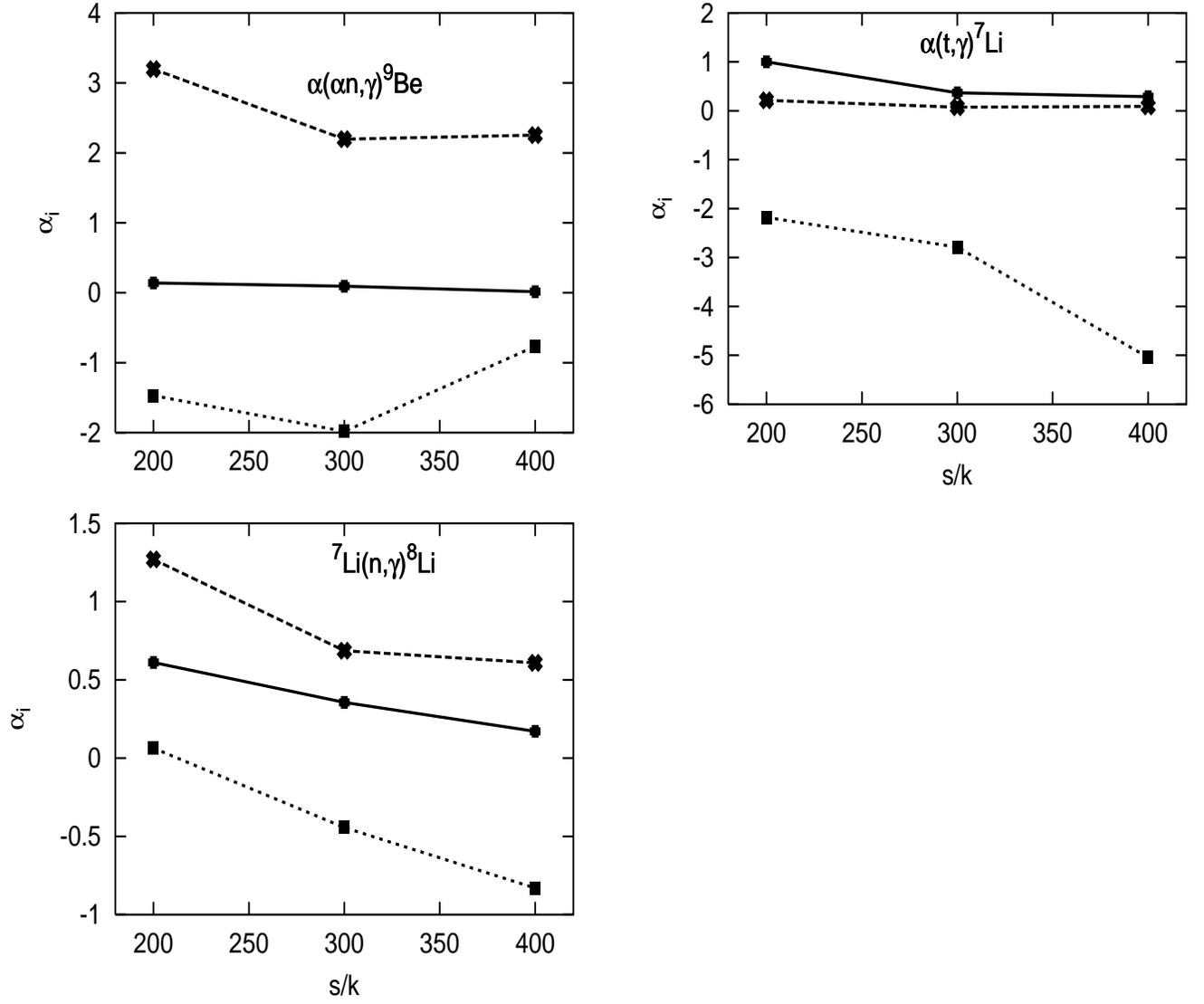}}
\caption{Dependence of the sensitivity parameter, $\alpha_{i}$, on the entropy per
 baryon, $s$/k, corresponding to the parameter sets in Table 
 \ref{flowparameterentro}. Three symbols are the same as those in Fig.\ref{fgtau_nama}.  
 The dependence on $s$/k is very small, except for actinide elements (squares).}
\label{fgentropy}
\end{figure}

\begin{figure}
\centering
\rotatebox{0}{\includegraphics[width=18cm,height=15cm]{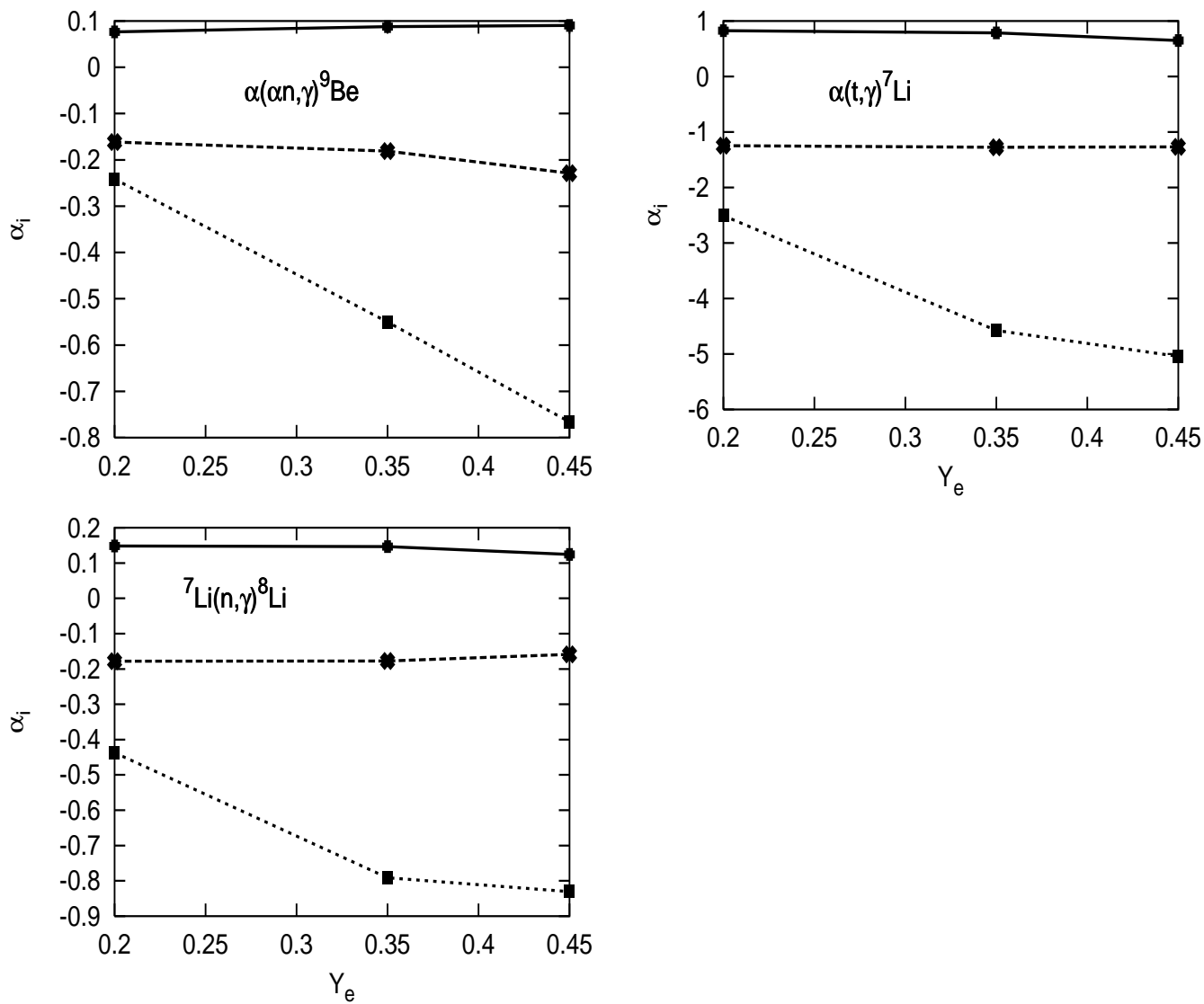}}
\caption{Dependence of the sensitivity parameter, $\alpha_{i}$, on the
 initial electron fraction, $Y_{e}$, corresponding to the parameter sets given in Table
 \ref{flowparameterye}. Three symbols are the same as those in
 Fig.\ref{fgtau_nama}. The dependence on $Y_{e}$ is very small, except
 for actinide elements (squares).}
\label{fgye_nama}
\end{figure}

\begin{figure}
\centering
\rotatebox{0}{\includegraphics[width=18cm,height=15cm]{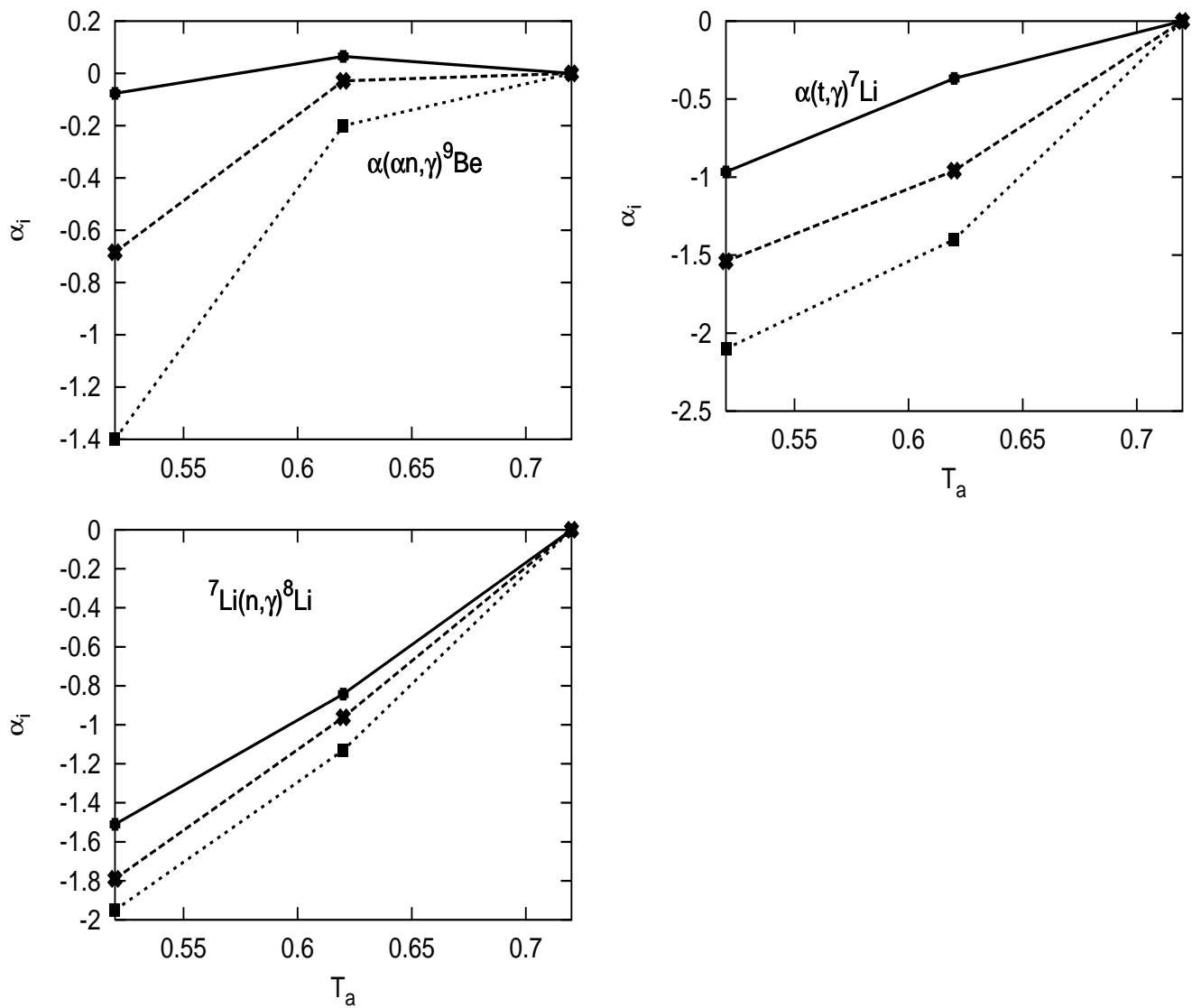}}
\caption{Dependence of the sensitivity parameter, $\alpha_{i}$, on the
 asymptotic temperature, $T_{a}$, corresponding to the parameter sets given in Table 
 \ref{flowparametertemp}. Three symbols are the same as those in Fig.\ref{fgtau_nama}. The dependence on $T_{a}$ is strong, but
 weaker than the dependence on $\tau_{dyn}$ (see
 Fig. \ref{fgtau_nama}).} 

\label{fgtemp}
\end{figure}

Figures \ref{fgtau_nama} - \ref{fgtemp} 
illustrate the dependence of the sensitivities 
of the three most important reactions (1)-(3) on
various characteristic quantities
based upon the parameter sets shown in Tables
\ref{flowparametertau}
- \ref{flowparametertemp}.  
In extreme conditions which have a wide range of $\tau_{dyn}$ and even
very low $Y_{e}$ such as in the prompt SN explosions or neutron-star
mergers, we need to know the reaction sensitivities for more nuclear
reactions in Table \ref{reaction}. 
We therefore summarize in Tables \ref{resulta} -
\ref{resultd} the detailed sensitivity results of the extended reactions
(1)-(6) and (15) for each model.
From these results we deduce
that the dynamical timescale ${\tau}_{dyn}$ and asymptotic temperature
$T_{a}$ can significantly affect the sensitivity, while $s/k$ and $Y_{\rm{e}}$ 
do so only moderately. Note the difference of vertical scale for
$\alpha_{i}$ in Figure \ref{fgentropy} ($\tau_{dyn}$ = 1.0 ms) and Figure  
\ref{fgye_nama} ($\tau_{dyn}$ = 5.0 ms). 
Although the sensitivity does not strongly change for various $s/k$ or
$Y_{e}$, absolute $\alpha_{i}$ values are generally larger for shorter
dynamical timescale. 

These indicate that knowing ${\tau}_{dyn}$ for the true r-process
environment is most crucial for
determining the reaction sensitivities.
Indeed, ${\tau}_{dyn}$ determines    
whether the $\alpha$-process dominates the
neutron-capture process, 
and is thus the key parameter 
to understanding the r-process in supernova heated
winds. 
The second most important parameter is $T_{a}$. 
This parameter determines how long the neutron-capture flow endures before
freezeout. 

However, ${\tau}_{dyn}$ and $T_{a}$ 
can only be estimated from 
a realistic, definitive model for SN
explosions which does not yet exist. 
Moreover, actinide elements are the most
sensitive to all astrophysical environmental parameters (Otsuki et
al. 2003; Sasaqui, Kajino \& Balantekin 2005), as clearly shown in
Figures \ref{fgtau_nama} - \ref{fgtemp} and Tables \ref{resulta} - \ref{resultd}, because a slight
variation 
of the nuclear reaction rate results in a large effect on the yield,
as shown in Fig \ref{ys}.  This makes determining of 
accurate light-element reaction rates crucial for cosmochronology.

\section{Detailed Analysis of the Nuclear Reaction Flow}
\label{carbon}

\subsection{Carbon Isotopes; Neutron-capture vs. $\alpha$-capture}
\begin{figure}
\centering
\rotatebox{0}{\includegraphics[width=15cm,height=13cm]{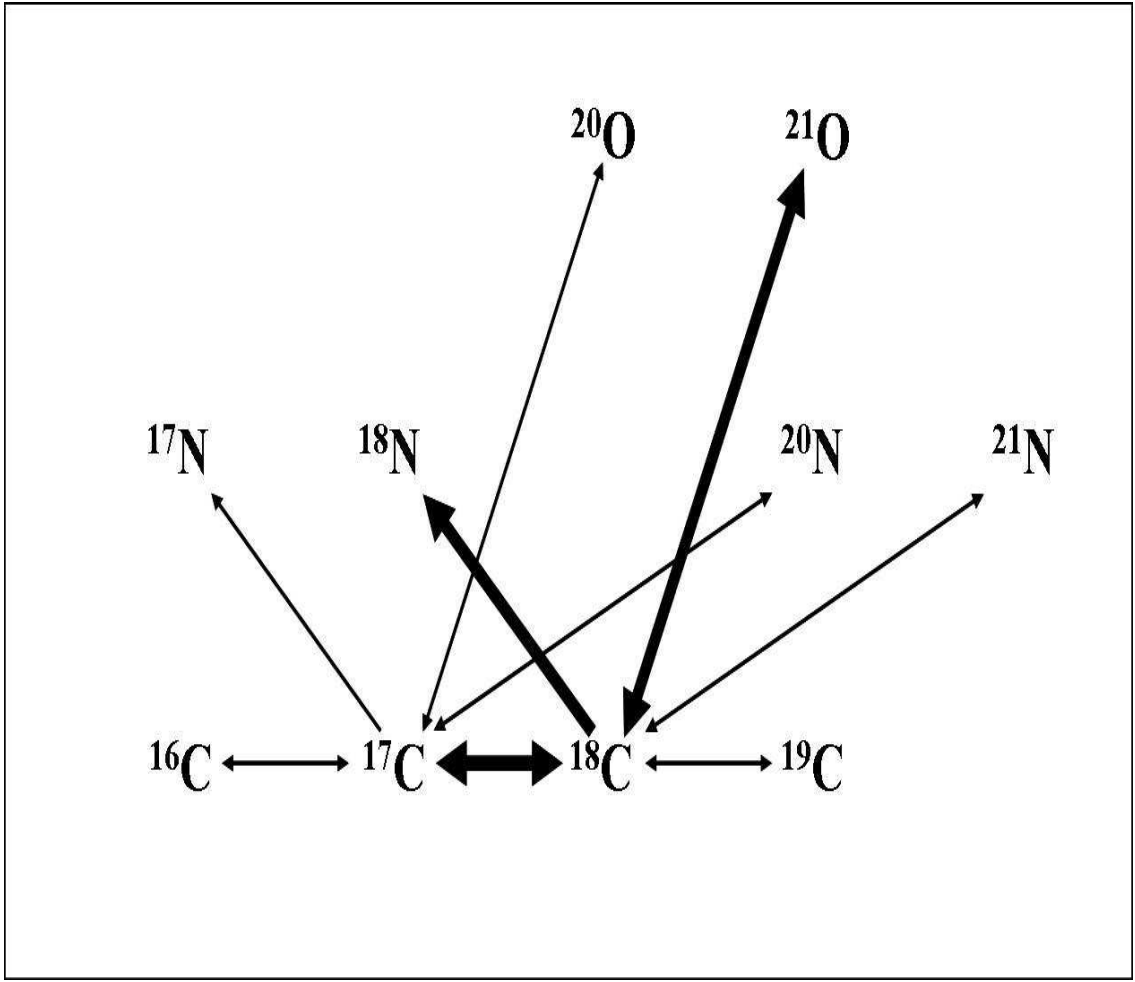}}
\caption{The reaction flows near $^{18}$C. 
 The thick arrows indicate the most 
 important reactions, which lead to finite sensitivities in the
 r-process nucleosynthesis shown in Tables \ref{result1} - \ref{resultd}.
}
\label{f4}
\end{figure}

We found in Sect.5 that many carbon isotopes are on the most important
nuclear reaction chains. We therefore study the importance of these
nuclei in the r-process 
nucleosynthesis more in detail in this section.  
There are four important flow paths: They 
are the 
neutron-capture, the photo-disintegration,
the $\alpha$-capture and the $\beta$-decay reactions (Figure \ref{f4}).

The time evolution of
the abundances is given,
\begin{eqnarray}
{{{\rm d}\mathrm{Y}_{{}^{i}{\mathrm{C}}}}\over{{\rm d}{\mathrm{t}}}}&=&
-N_{A}\langle{\sigma
v_{{}^{i}{\mathrm{C}}(\mathrm{n},{\gamma})}}\rangle{\rho}\mathrm{Y}_{{}^{i}{\mathrm{C}}}\mathrm{Y}_{\mathrm{n}}
+{\lambda_{i+1}}\mathrm{Y}_{{}^{i+1}{\mathrm{C}}}\nonumber\\
&&-{\lambda_{i}}\mathrm{Y}_{{}^{i}{\mathrm{C}}}
+N_{A}\langle{\sigma
v_{{}^{i-1}{\mathrm{C}}(\mathrm{n},{\gamma})}}\rangle{\rho}\mathrm{Y}_{{}^{i-1}{\mathrm{C}}}\mathrm{Y}_{\mathrm{n}}\nonumber\\
&&-N_{A}\langle{\sigma
v_{{}^{i}{\mathrm{C}}({\alpha},\mathrm{n})}}\rangle{\rho}\mathrm{Y}_{{}^{i}{\mathrm{C}}}\mathrm{Y}_{\alpha}
+N_{A}\langle{\sigma
v_{{}^{i+3}O(\mathrm{n},{\alpha})}}\rangle{\rho}\mathrm{Y}_{{}^{i+3}\mathrm{O}}\mathrm{Y}_{\mathrm{n}}\nonumber\\
&&-N_{A}\langle{\sigma
v_{{}^{i}{\mathrm{C}}({\alpha},p)}}\rangle{\rho}\mathrm{Y}_{{}^{i}{\mathrm{
C}}}\mathrm{Y}_{\alpha}
+N_{A}\langle{\sigma
v_{{}^{i+3}{\mathrm{N}}(p,{\alpha})}}\rangle{\rho}\mathrm{Y}_{p}\nonumber\\
&&-{\lambda_{\beta}}\mathrm{Y}_{{}^{i}\mathrm{C}}~~, 
\end{eqnarray}
where \(\mathrm{Y}_{{}^{i}{\mathrm{C}}}\) is the number abundance of a C
isotope with mass number i, while 
\(\mathrm{Y}_{{}^{i+1}{\mathrm{C}}}\),
\(\mathrm{Y}_{{}^{i-1}{\mathrm{C}}}\),
\(\mathrm{Y}_{{}^{i+3}N}\) and
\(\mathrm{Y}_{{}^{i+3}O}\) represent the nuclei made by
$(\rm{n},\gamma)$, $(\gamma,\rm{n})$, $(\alpha,p)$ and
$(\alpha,\rm{n})$ reactions, respectively, and $\lambda_{i}$ is the
photo-neutron emission rate for $^{i}$C($\gamma$,n)$^{i-1}$C which
should satisfy the principle of the detailed balance with the forward
(n,$\gamma$) rate. 
The most important paths are ($\alpha$,n), (n,$\gamma$), and
$\beta$-decay reactions, as shown by thick solid arrows in Figure \ref{f4}.

\begin{figure}
\centering
\rotatebox{0}{\includegraphics[width=15cm,height=13cm]{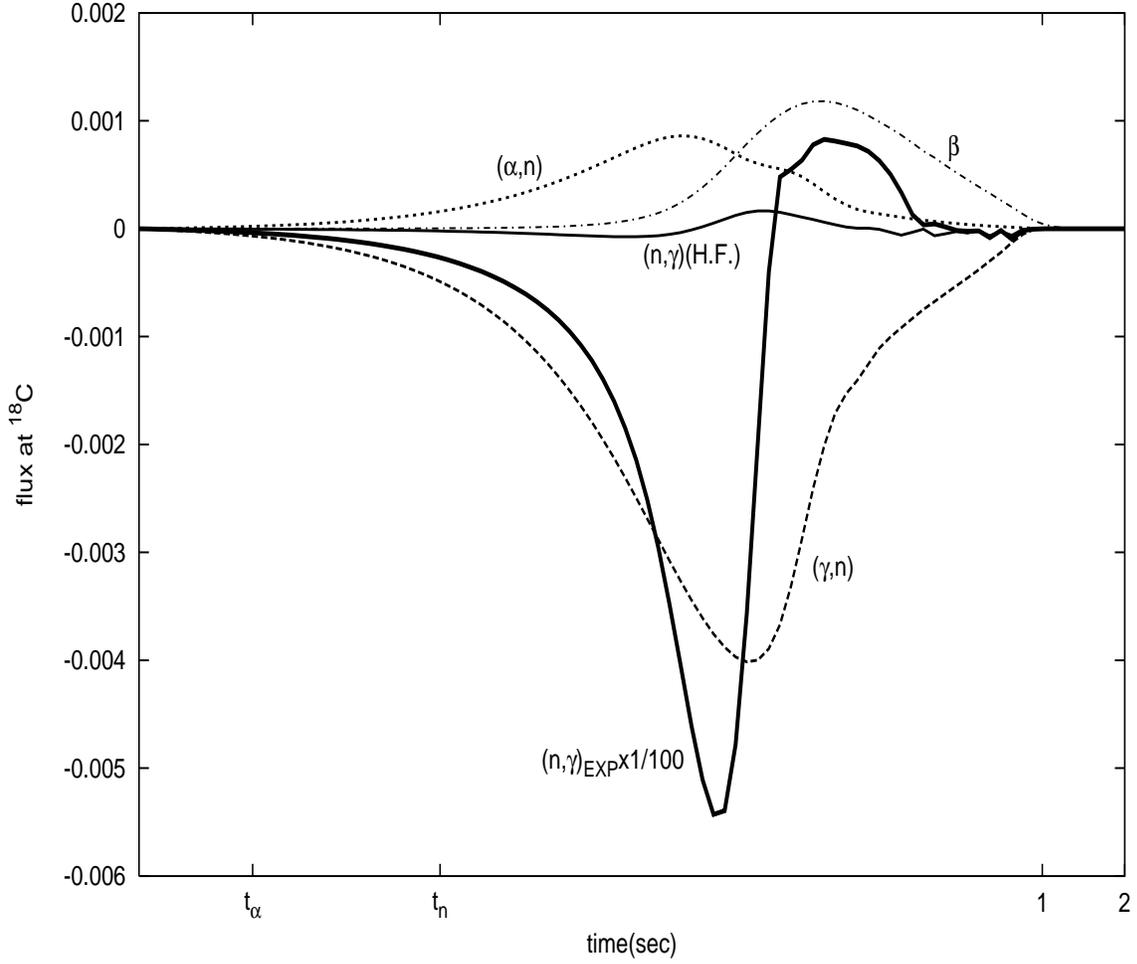}}
\caption{Net nuclear flux through ${}^{18} \mathrm{C}$ as a function of
 time, Eq.(16).  
Positive values denote flux
 out of ${}^{18} \mathrm{C}$, while negative values denote flux flowing 
 into ${}^{18} \mathrm{C}$.
 $t_{\alpha}$ indicates the time at which NSE breaks down and the
 $\alpha$-process effectively begins.
 $t_{\mathrm{n}}$ is the time at which the $\alpha$-process freezes out and
 n-captures begin.
 The solid lines show the $^{18}$C$(\mathrm{n},\gamma)^{19}$C flow when the
 HF estimate (thin solid line) or  the experimental data (thick solid line) were used as labeled.
 The dotted line, the dot-dash line, and the dash line express the 
 $^{18}$C$(\alpha,\mathrm{n})$,  $\beta$-decay, and
 $^{18}$C$(\gamma,\mathrm{n})$ flows, respectively.
 Note that the $(\mathrm{n},\gamma)$ and $(\gamma,\mathrm{n})$ 
 rates always dominate over the 
 $(\alpha,\mathrm{n})$ and $\beta$-decay rates.
 With the HF estimates, however,  the n-capture reactions are ineffective, so that 
 $\alpha$-capture and $\beta$-decay processes balance the
 photo-disintegration. With the experimental rate, n-capture
 dominates the other three reactions. Even so, most of the flux
 flows to ${}^{18} \mathrm{C}$ so that it becomes a new 
 ``semi-waiting'' point (see text in Sect. 6.2 in detail).}
\label{fg58}
\end{figure}

If the neutron separation energy of $^{i+1}$C is large, 
the (n,$\gamma$) capture of $^{i}$C into the isotope $^{i+1}$C exceeds its 
($\gamma$,n) photodisintegration. This is the case for $^{18}$C, and we
show in Figure \ref{fg58} the net nuclear flux $F_{ij}$(t) which in
defined by Eq.(16). This results in an 
accumulated abundance of carbon nuclei with the highest neutron
separation energies 
(e.g. $^{16}$C and $^{18}$C in Table \ref{tablecarbon}).
It is the reason for the large sensitivity of the r-process to the
${}^{15} \mathrm{C}$(n,$\gamma$)${}^{16} \mathrm{C}$ and
${}^{17} \mathrm{C}$(n,$\gamma$)${}^{18} \mathrm{C}$ reactions (see Table \ref{result1}).

On the other hand, 
the $\alpha$-captures occur until just before the time $t_{\mathrm{n}}$
at which the $\alpha$-process
freezes out and the neutron-capture process effectively starts.
The $(\alpha, \rm{n})$ reaction in particular, can be as important
 as the
$(\rm{n},\gamma)$ or the inverse reactions (for example, see Figure
\ref{fg58} for $^{18}$C). 
We will describe the ${}^{18}\mathrm{C}$ in detail in Section 6.2.
In general, $\beta$-decay is not important for the abundance evolution 
of long lived ($t_{1/2} > 1 s$) C isotopes  (Table \ref{tablecarbon}).
The main reactions for C isotopes are $(\mathrm{n},\gamma)$
and $(\gamma,\mathrm{n})$, while the $\beta$-decay and $\alpha$-capture
on $^{18} \rm{C}$ operate as an escape route to heavier elements.

These trends also appear in the evolution of
other carbon nuclei. For example, 
 the reaction flux between ${}^{14} \mathrm{C}$ and ${}^{15}
\mathrm{C}$,  ${}^{16} \mathrm{C}$ and ${}^{17} \mathrm{C}$, 
and  ${}^{18} \mathrm{C}$ and ${}^{19} \mathrm{C}$
is similar. 
Just after $t_{\mathrm{n}}$,
 photo-disintegration dominates at
${}^{15} \mathrm{C}$, ${}^{17} \mathrm{C}$ and  ${}^{19} \mathrm{C}$,
because of their small neutron separation energies $\mathrm{S_{n}}$
(see Table \ref{tablecarbon}).
However, as the 
temperature drops  the neutron-captures
proceed efficiently with no $\alpha$-captures.
The direction of the flux is thus toward heavier nuclei.
On the other hand, between 
${}^{15} \mathrm{C}$ and ${}^{16} \mathrm{C}$, 
or between ${}^{17} \mathrm{C}$ and ${}^{18}\mathrm{C}$,
the net flux is always toward  heavy isotopes by neutron-capture. 

\subsection{$^{18}$C(n,$\gamma$)$^{19}$C and the $^{18}$C Waiting Point}

\begin{figure}
\centering
\rotatebox{0}{\includegraphics[width=18cm,height=15cm]{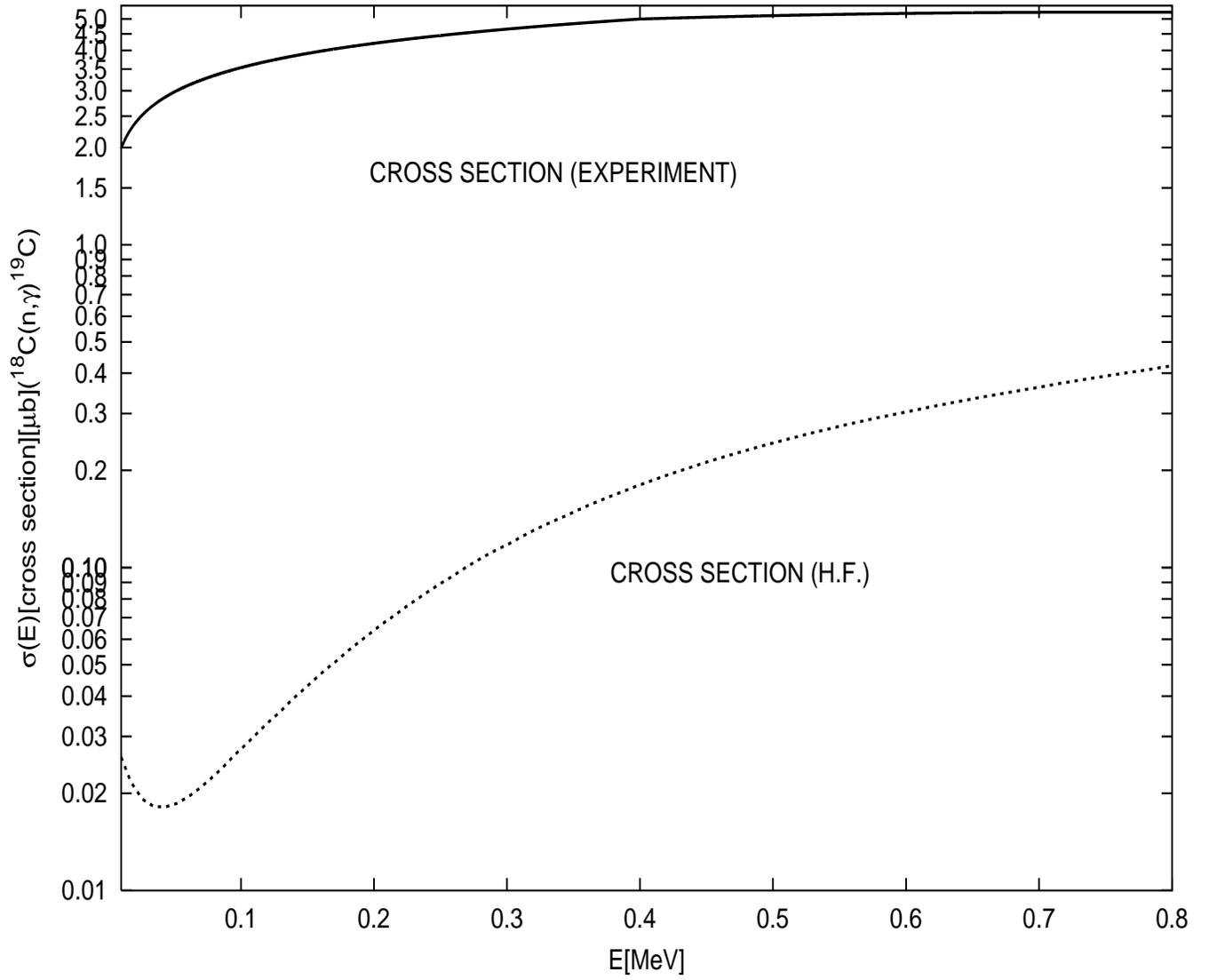}}
 \caption{Comparison of the experimentally 
 determined $^{18}$C($\mathrm{n}$,$\gamma$)$^{19}$C
 cross section  (solid line) of Nakamura et al. (1999) with
  Hanser-Feshbach estimates (dotted line).
 }
\label{f3}
\end{figure}

\begin{figure}
\centering
\rotatebox{0}{\includegraphics[width=15cm,height=13cm]{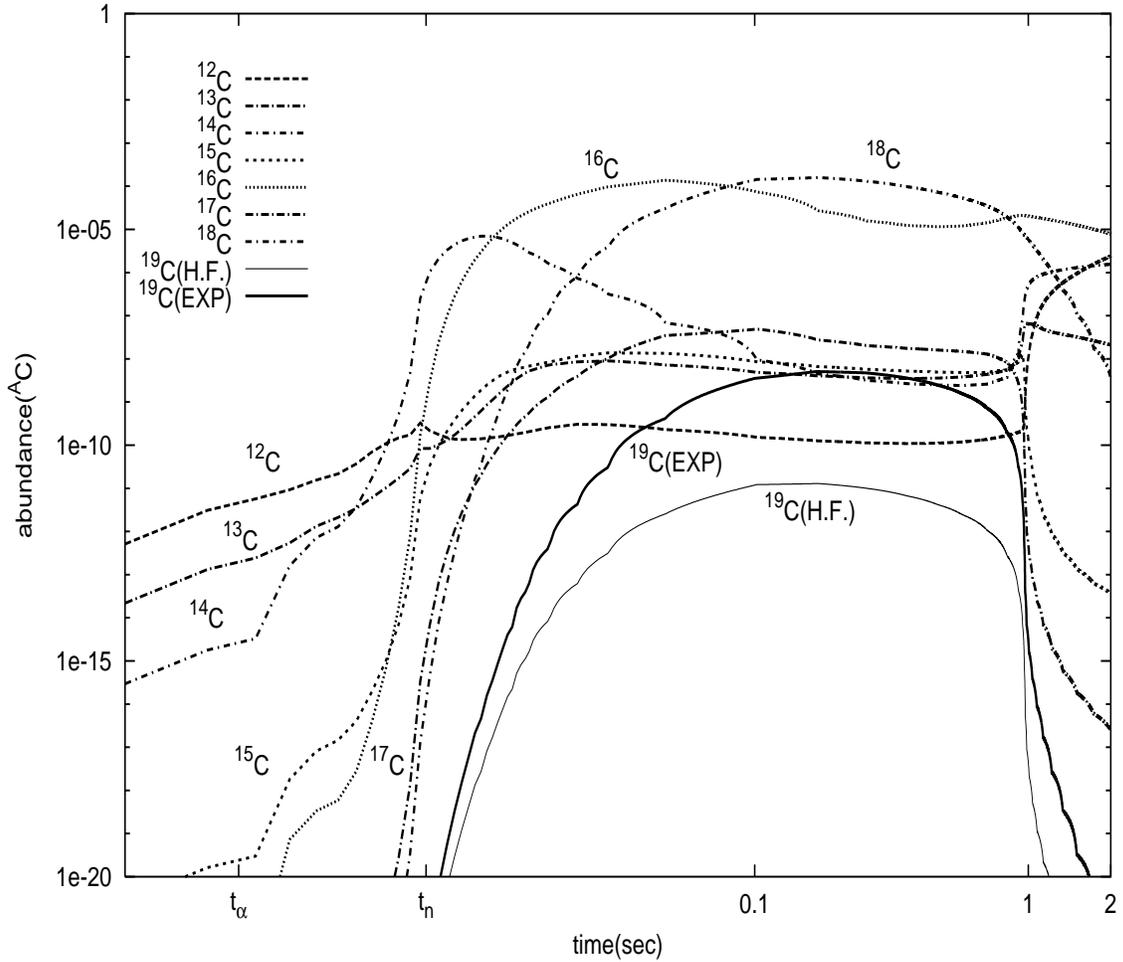}}
\caption{Time evolution of carbon isotope abundances. Because of the large
 measured reaction
 cross section for $^{18}$C(n,$\gamma$)$^{19}$C, the abundance of 
 $^{19}$C   is increased compared
 with the abundance based upon the HF estimate (see Figure \ref{f3}). However,
 the sensitivity of 
 this reaction to the r-process is small (see Table \ref{result1}). 
 The accumulated abundance of $^{19}$C rapidly 
 photodisintegrates to
 $^{18}$C due to the small neutron-separation energy (see Table
 \ref{flowparametertau}
 and text).}
\label{carbon_abundance}
\end{figure}

A discussion of the $^{18}$C(n,$\gamma$)$^{19}$C reaction rate is 
illuminating as an example of the importance of examining the
reaction flow when determining which reactions are most crucial 
to measure.   It also illustrates the uncertainties in
the nuclear reaction rates in this region.  As noted above,
most reaction rates for these light nuclei have not
been measured.  A measurement of the cross section for the \({}^{18}\rm{C}(\mathrm{n},\gamma)
{}^{19}\rm{C}\) reaction (16) in Table \ref{reaction} was, however,
carried out by Nakamura et al. (1999) by using the Coulomb dissociation
method of $^{19}$C. Applying the principle of the detailed balance to
the ($\gamma$,n) photodistribution and the (n,$\gamma$)
neutron-capture, we can obtain the forward (n,$\gamma$) cross section
shown in Figure \ref{f3}.
This Figure \ref{f3} compares
cross sections from the HF model and the experiment.
The experimental data are about two orders of magnitude larger than
the theoretical HF estimates.
The experiment by Nakamura et al.(1999) also shows that the
neutron separation energy of 
${}^{19} \mathrm{C}$ is 0.530 MeV which is larger than the
theoretical estimate of 
0.191 MeV. 
These two measurements 
lead to a drastic increase of the abundance of 
$^{19}$C (thick solid line in Figure \ref{carbon_abundance}) relative
to that based upon the HF cross sections (thin solid line in Figure
\ref{carbon_abundance}).  

On the other hand, even with this drastic change in the reaction rate,
there is no sensitivity of the r-process to this reaction $i.e.$
$\alpha_i \approx 0$. (See Table \ref{result1}.)
The abundance of $^{18}\rm{C}$ remains
almost unchanged (see Figure \ref{carbon_abundance}). 
In other words, the nuclear flow through $^{18}$C is the same regardless of the 
(n,$\gamma$) cross section. This 
is explained by the fact that 
the ($\gamma$,n) rate is faster than the (n,$\gamma$).
The neutron separation
energy between $^{18}$C and $^{19}$C is ``small'' compared with other
carbon nuclei even though it was increased from 0.191 MeV to 0.53 MeV. 
 (See Table \ref{tablecarbon}). 
Hence, $^{19}\rm{C}$ is quickly photo-disintegrated.
The reaction path then  flows out into heavy elements through $^{18}$C($\alpha$,n)$^{21}$O and even 
though the (n,$\gamma$) rate indirectly affects the r-process, 
there is essentially no sensitivity of the r-process to this reaction. 

\begin{figure}
\centering
\rotatebox{0}{\includegraphics[width=18cm,height=15cm]{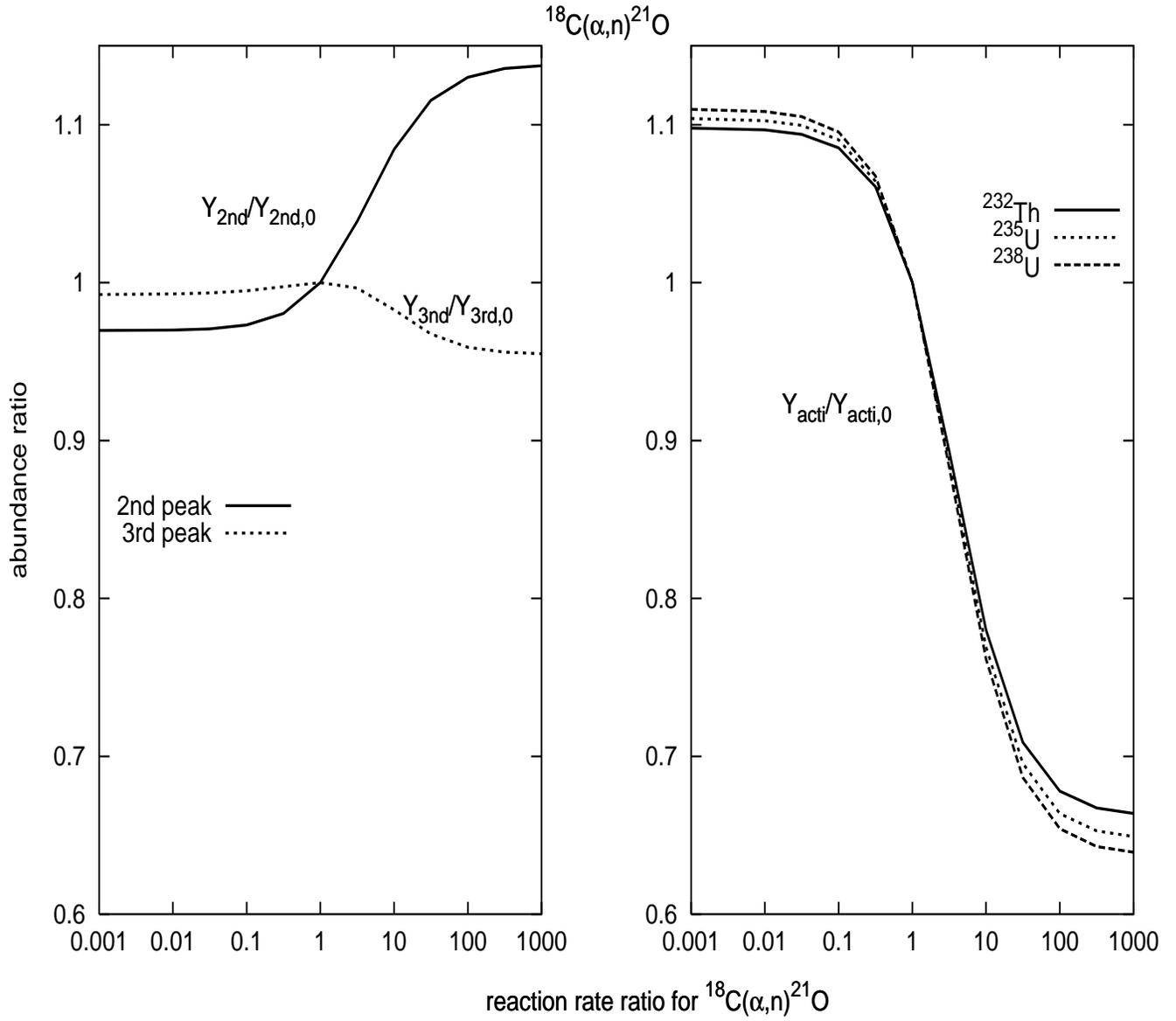}}
\caption{Calculated abundance ratios for the 2nd or
 3rd peak (left) and actinide (right) elements as a function of the
 reaction rate \(\lambda_{i}/\lambda_{i}(0)\) for   
 $^{18}$C($\alpha$,n)$^{21}$O.}
\label{fgra}
\end{figure}

It is also of interest that 
the reaction flow stalls for an instant at $^{18}\rm{C}$.
The competition between $(\alpha,\rm{n})$ and
$\beta$-decay is then important.
An experimental measurement of the $^{18}$C($\alpha$,n)$^{21}$O
reaction has not yet been completed.  However,
if obtained the new value might well be very different 
as it was for the $^{18}\mathrm{C}(\mathrm{n},\gamma)^{19}\mathrm{C}$ reaction.
Hence, we have tested a large range for this ($\alpha$,n) reaction rate. Figure \ref{fgra}
shows the produced abundance sensitivity
to the cross section ratio about the 2nd or 3rd peaks (left panel
in Figure \ref{fgra}) and
actinides (right panel in Figure \ref{fgra}). 
If the cross section increases by a factor of 
100,
the abundances of the 2nd and 3rd peaks are estimated to change by about
10$\%$, while the actinides would change by about 30$\%$.

\begin{figure}
\centering
\rotatebox{0}{\includegraphics[width=14cm,height=17cm]{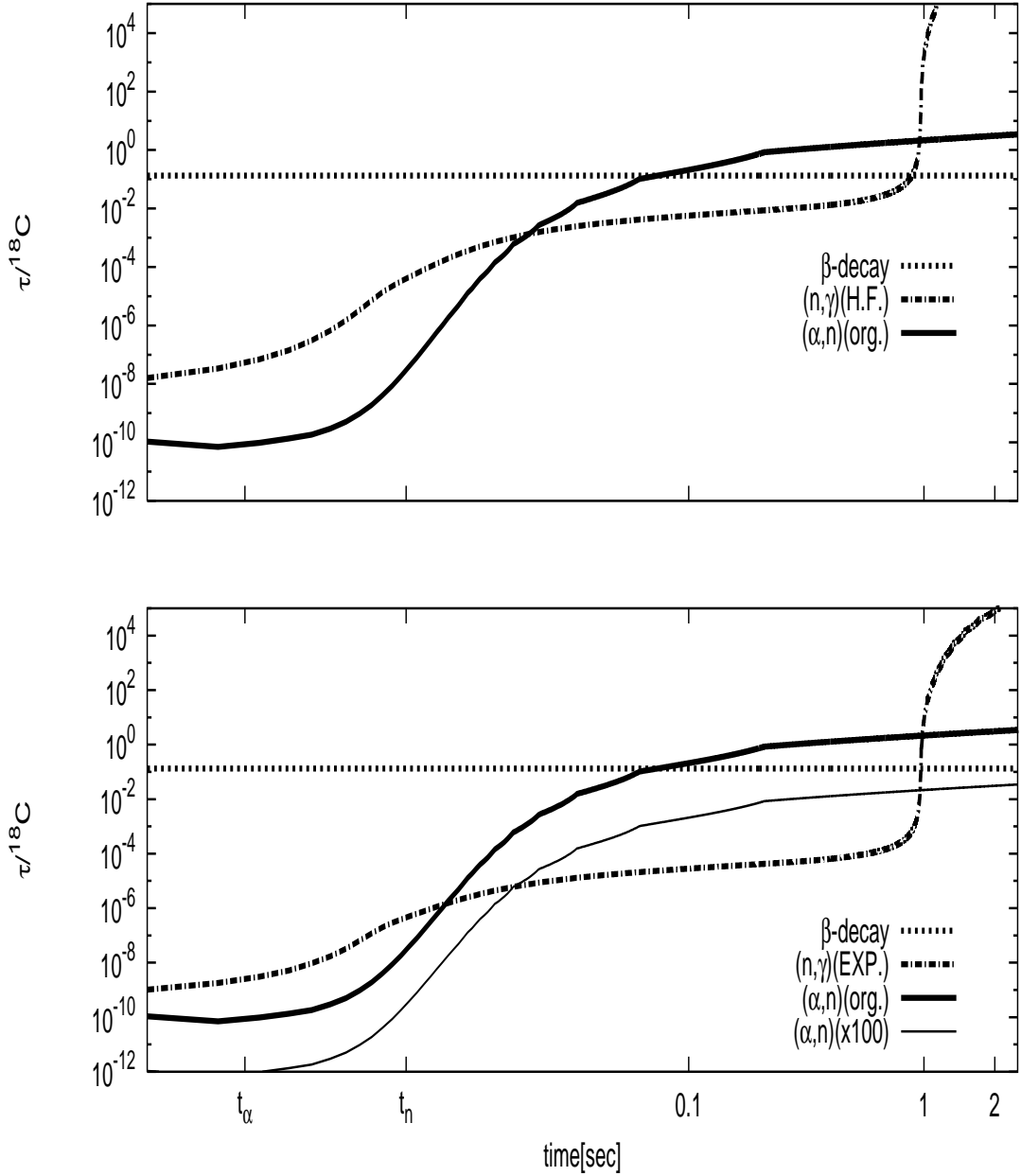}}
\caption{Reaction time scales per $^{18}$C for (n,$\gamma$),
 $(\alpha,\rm{n})$ and 
 $\beta$-decay. The top panel shows the result based upon HF estimates
 for the (n,$\gamma$) reaction rate. 
 The lower panel shows the result of using the experimental (n,$\gamma$)
 rate. For illustration, the lower panel also shows the result
 of changing the ($\alpha$,n) cross section by a factor of 100.
 Increasing the $(\alpha,\rm{n})$ cross section leads to a 
 change of the order of the reaction flow speed through
 $(\alpha,\rm{n})$. 
 $t_{\alpha}$ and $t_{n}$ are the same as those in Figure \ref{fg58}.}
\label{fgspeed}
\end{figure}

In the range of $\lambda/\lambda_{0}=100$, the sensitivity of the 
$(\alpha,\rm{n})$ reaction is significant. On the other hand,
in the range of $\lambda/\lambda_{0}=[100-1000]$, the estimated abundance
is almost
constant (right side of Figure \ref{fgra}). In this region
it has no sensitivity. 
This is because 
when the cross section for the $(\alpha,\rm{n})$ reaction becomes larger, 
it exceeds the rate for  $\beta$-decay 
$\lambda_{^{18}\rm{C}(\alpha,\rm{n})^{21}\rm{O}}/\lambda_{\beta}\geq 10$
throughout the nucleosynthesis process (see the lower
panel of Figure \ref{fgspeed}). In this region, 
the $(\alpha,\rm{n})$ reaction plays a very important role
because it determines whether the reaction flows 
through $(\alpha,\rm{n})$ or $\beta$-decay. 
However, in the range $\lambda/\lambda_{0}=[100-1000]$,
the order of the reaction flow speed between the $(n,\gamma)$,
the $(\alpha,\rm{n})$ and the $\beta$-decay remains unchanged.
The reaction flux stays at $^{18}\rm{C}$ and only waits for the $\beta$-decay
because the reaction flow of both $(n,\gamma)$ and $(\alpha,\rm{n})$
reactions are much faster than $\beta$-decay. In essence, $^{18}\rm{C}$ becomes
a ``semi-waiting point''. 
As a result, the reaction flow follows only one path that is
$^{17}$C$\rightarrow ^{18}$C$\rightarrow ^{18}$N.
Therefore the $(\alpha,\rm{n})$ rate does not affect the r-process.
Hence, we find that the reaction
$^{18}\mathrm{C}(\alpha,\mathrm{n})^{21}\mathrm{O}$ 
is most important for the r-process among all reactions on $^{18}$C. 



\section{Summary and Conclusion}
In this work, we have quantified the uncertainty of r-process
nucleosynthesis to various light-element nuclear reactions. 
We checked the dependence of the sensitivities for 
several parameter sets in explosive environments ($\tau_{dyn}$, $s/k$, $Y_{e}$
and $T_{a}$).
We found that knowing the dynamical timescale ${\tau}_{dyn}$ is most crucial for determining the reaction sensitivity for the light-element reactions studied here.

We identified the important reactions for the r-process by introducing the
concept of sensitivity. The \(\alpha (\alpha \mathrm{n}, \gamma)
^{9}\mathrm{Be}\) reaction is the most important reaction regardless of the flow
models, as has been noted previously. In the present work, however,  we
also identify other equally important
reactions for synthesizing the r-process elements in SNe. 
They are primarily the \(\alpha (\mathrm{t}, \gamma)^{7}\mathrm{Li}\) and 
\(^{7}\mathrm{Li}(\mathrm{n}, \gamma) ^{8}\mathrm{Li}\) reactions, 
though other relatively important reactions were also found. 
We also carried out detailed analysis of the sensitivity for all
other eighteen nuclear reactions by comparing the calculated results
with the use of new experimental data
and H.F. estimates. 
We have also analyzed the 
semi-waiting point at $^{18}$C.   An interesting  feature of light-element
waiting points is the importance of the competition between
$\alpha$-capture, neutron capture, and $\beta$-decay in the 
nuclear reaction flows for r-process
nucleosynthesis. This applies generally to other light-mass nuclei such as
$^{8}$Li, $^{15}$B, $^{16,18}$C, and $^{24}$O,  
whose reaction Q-values or neutron capture rates are small. 
The only difference from the well studied waiting-point nuclei in
heavier-mass regions (like $^{130}$Cd and $^{132}$Sn) is that 
$\alpha$-capture is an additionally important flow which competes with
neutron capture and $\beta$-decay in the light-mass region. This specific
feature arises from the fact that the $\alpha$-process, which proceeds
prior to the neutron-capture flow on the intermediate-to-heavy mass
nuclei, occurs at early times in the  high temperature and high density of
various expansion of the neutrino-driven wind. 

We also studied extensively the dependence of the sensitivity on the
expansion flow models.  We adopted 
exponential flow models to simulate various expansion dynamics and
calculated the sensitivity. The four important physical
parameters which characterize the profile in the flow models are 
the dynamical timescale $\tau_{dyn}$, the entropy per baryon $s/k$, the initial electron
fraction $Y_{e}$, and the asymptotic temperature $T_{a}$.   
We found that  the
sensitivities depend most strongly  on the dynamical timescale. 

Nuclear reactions and SN dynamics are thus
complementary facets to 
understanding r-process nucleosynthesis. 
Obviously, more  accurate experimental data should help 
our comprehensive understanding of r-process nucleosynthesis and SN
dynamics.
We believe that the present work provides a useful tool to elucidate 
which reactions are most important and
also the synergy between the SN dynamics and 
nuclear reactions in r-process nucleosynthesis.

\acknowledgments
This work has been supported in part by Grants-in-Aid
for Scientific Research (13640313) and
for Specially Promoted Research (13002001) of the Ministry of Education,
Science, Sports and Culture of Japan, and The Mitsubishi Foundation.
Work at The University of Notre Dame has been 
supported under DoE nuclear theory grant DE-FG02-95-ER 40934.


\section{Appendix A}
\subsection{Justification of the Incoherent Approximation for $\alpha_{i}$}
\label{justif}

We have discussed the reaction sensitivity by making the incoherent 
approximation, Eq.~(\ref{approx2}), assuming that $\alpha_{i}$ is most sensitive to the
change of the corresponding i-th nuclear reaction rate $\lambda_{i}$ and
that its
dependence on different $\lambda_{j}$'s ($j \ne i$) is weak. 
In order to justify 
this approximation quantitatively, we carried out several
calculations of $\alpha_{i}$ by changing two or three reaction rates
\{$\lambda_{j}$\} simultaneously. For this purpose we chose the three most
important nuclear reactions from Tables \ref{result1} and \ref{result2}, i.e. (1)
$\alpha(\alpha \rm{n}, \gamma)^9 \rm{Be}$, (2) $\alpha(\rm{t}, \gamma)^7
\rm{Li}$, and (3) $^7 \rm{Li}(\rm{n},\gamma)^8 \rm{Li}$. These reactions
show fairly large absolute values of the sensitivity parameter
$\alpha_{i}$.  

\begin{figure}
\centering
\rotatebox{0}{\includegraphics[width=18cm,height=15cm]{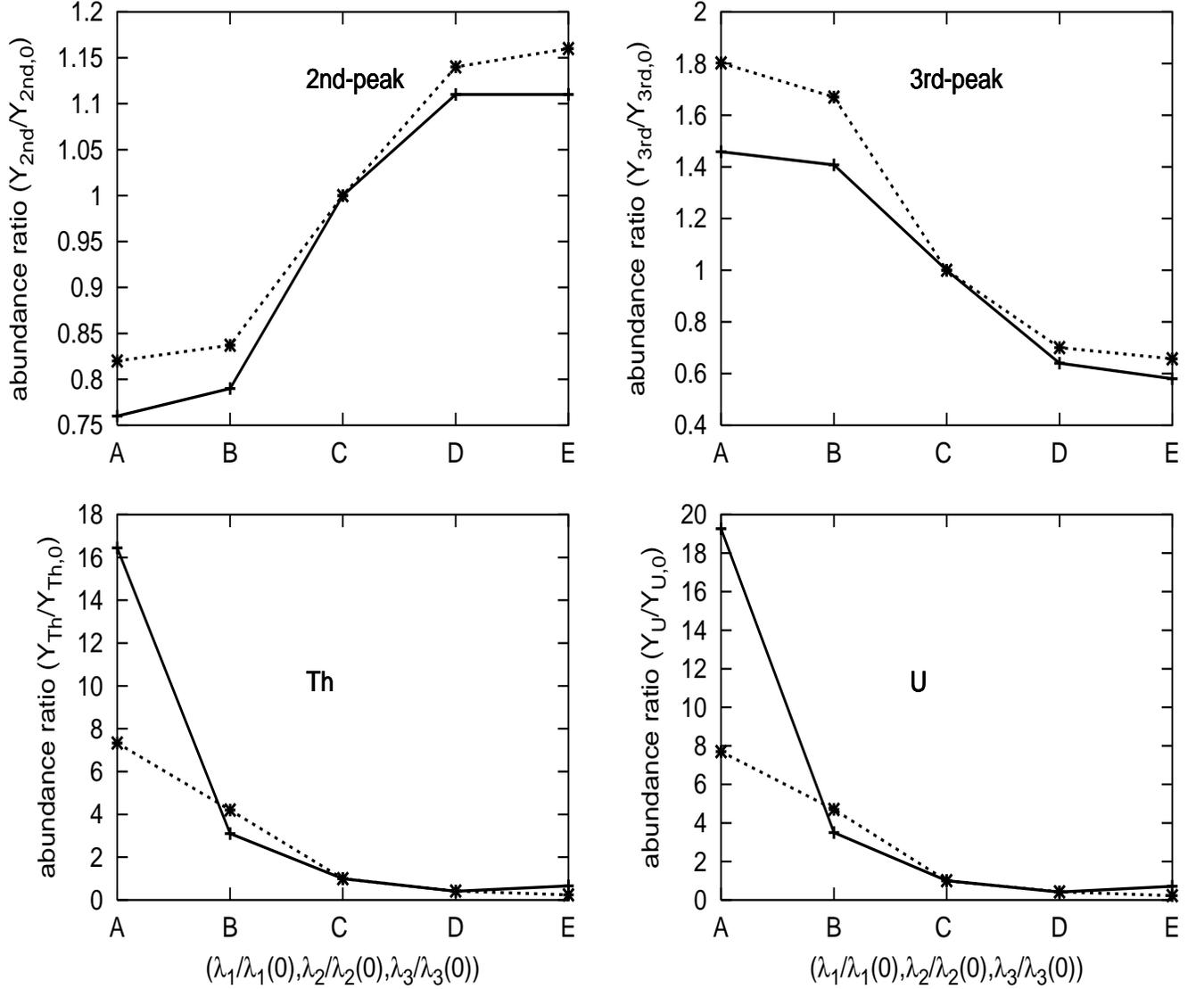}}
\caption{Calculated r-process yields for the triple variation of
 reaction rates.
 Solid curves show the exact calculation, 
 while the  dotted curves are based upon  
 the incoherent approximation, Eq.(7). 
 The points 'A', 'B', 'C', 'D', and 'E' corresponds to
 ($\lambda_{1}/{\lambda_{1}(0)}$, $\lambda_{2}/{\lambda_{2}(0)}$,
 $\lambda_{3}/{\lambda_{3}(0)}$)=(0.65,0.70,0.65),
 (0.65,0.70,1.0), (1.0,1.0,1.0), (1.35,1.30,1.0), and (1.35,1.30,1.35),
 respectively.}
\label{f2}
\end{figure}

Figure \ref{f2} displays the calculated r-process yields, in which
we have changed
the three reaction rates as indicated. 
Solid curves show the results of the exact calculation 
and the dotted curves show the results of 
making the incoherent approximation 
\begin{eqnarray}
{{Y_{r}}\over{Y_{r}(0)}}\simeq \Biggl({{\lambda_{1}}\over{\lambda_{1}(0)}}\Biggr)^{\alpha_{1}}\Biggl({{\lambda_{2}}\over{\lambda_{2}(0)}}\Biggr)^{\alpha_{2}}\Biggl({{\lambda_{3}}\over{\lambda_{3}(0)}}\Biggr)^{\alpha_{3}},
\end{eqnarray}
where the $\alpha_{i}$'s are taken from Table \ref{result1} for the fast wind model.
The exact (solid) and approximate (dotted) calculations are in
reasonable agreement with each other for the r-process abundances
in the 2nd and 3rd peaks. A large deviation, however, emerges in actinide
elements when the three reaction rates are set to near the lower limits
of  the estimated uncertainties of the cross sections. 

These results indicate that the incoherent approximation is a fairly 
good approximation, except for the actinide elements. We attribute this
deviation for actinide nuclei to the fact that the three most important
reactions change the neutron-to-seed ratio in slightly different manners. This
enhances the sensitivity in the actinide production near the end
of the neutron-capture flow along the r-process path. (See Table \ref{result1} and
Otsuki et al. 2003.)

\clearpage

\clearpage

\clearpage

\clearpage

\clearpage

\clearpage

\clearpage

\clearpage

\clearpage

\clearpage

\clearpage

\clearpage

\clearpage

\clearpage

%

\begin{table}
\raggedright
\caption{The most important 18 light-mass nuclear reactions, adopted 'standard'
 thermonuclear reaction rates, $\lambda_{i}(0)$, and uncertainties.}
\makebox[15.5cm]{\def\arraystretch{0.7}\begin{tabular}[t]{rllll}\hline
\multicolumn{2}{c}{reactions}&{\(N_{A}<\sigma v>\)}&1$\sigma^{b}$&{refs.$^{c}$}\\
\hline \hline
(1)&\(\alpha(\alpha\mathrm{n},\gamma){}^9\mathrm{Be}\)&
\(N_{A}^2<\alpha \alpha n>= 2.43\times 10^{9}T_{9}^{-2/3}\)&$\pm$35$\%$&
Sumiyoshi et al. 2002\\
&&\(\exp\{-13.490T_{9}^{-1/3}-(T_{9}/0.15)^2)\}\)&&\\
&&\(\times(1+74.5T_{9})
+6.09\times 10^5 T_{9}^{-3/2}\exp(-1.054/T_{9})\times\)&&\\
&&\([1-58.80T_{9}-1.794\times 10^{4}T_{9}^2+2.969\times 10^{6}T_{9}^3\)&&\\
&&\(-1.535\times 10^{8} T_{9}^4 +2.610\times 10^9 T_{9}^5 ]\)&&\\ \hline
(2)$^{a}$&\(\alpha(\mathrm{t},\gamma) {}^7\mathrm{Li}\)&
\(3.032\times 10^5 T_{9}^{-2/3}\exp(-8.09/T_{9}^{1/3})\)&$\pm$30$\%$
&Kajino et al. 1987\\
&&\([1.0 + 0.0516T_{9}^{1/3} + 0.0229T_{9}{2/3} + 8.28\times 10^{-3}T_{9}\)&&\\
&&\(-3.28\times 10^{-04}T_{9}^{4/3}-3.01\times 10^{-04}T_{9}^{5/3}]\)&&\\
&&\(+5.109\times
10^{5}T_{9\ast}^{5/6}T_{9}^{-3/2}\exp(-8.068/T_{9\ast}^{1/3})\)&&\\ \hline
(3)&\({}^7\mathrm{Li}(\mathrm{n},\gamma) {}^8\mathrm{Li}\)&\(
4.90\times 10^{3} + 9.96\times 10^{3} T_{9}^{-3/2}\exp(-2.62/T_{9})\)
&$\pm$35$\%$&Nagai et al. 1991b\\ \hline
(4)&\({}^8\mathrm{Li}(\alpha,\mathrm{n}) {}^{11}\mathrm{B}\)&
\(4.929\times 10^{6}T_{9}^{-3/2}\exp(-4.410/T_{9})\)
&$\times$ 2&X.Gu et al. 1995\\
&&\(+5.657\times 10^{8}T_{9}^{-3/2}\exp(-6.846/T_{9})\)
&\\
&&\(+4.817\times 10^{9}T_{9}^{-3/2}\exp(-11.836/T_{9})\)&&\\
&&\(+1.0\times10^{12}(10.03T_{9}^{-1}+4.814T_{9}^{-2/3})\)&&\\
&&\(\exp(-19.45/{T_{9}}^{1/3})\)&&\\ \hline
(5)&\({}^9\mathrm{Be}(\mathrm{n},\gamma) {}^{10}\mathrm{Be}\)&
\(1.01\times 10^{3}+1.01\times 10^{4}T_{9}^{-3/2}\exp(-6.487/T_{9})\)
&$\times$ 2&Rauscher et al. 1994\\
&&\(+5.41\times 10^{4}T_{9}^{-3/2}\exp(-8.471/T_{9})\)&&\\ \hline
(6)&\({}^{11}\mathrm{B}(\mathrm{n},\gamma) {}^{12}\mathrm{B}\)&
\(7.38\times 10^2 + 3.86\times 10^3T_{9}^{-3/2}\exp(-0.244/T_{9})\)
&$\times$ 2&Rauscher et al. 1994\\
&&\(+3.34\times 10^4 T_{9}^{-3/2}\exp(-4.99/T_{9})\)&&\\
\hline 
\label{reaction}
(7)&\({}^{12}\mathrm{B}(\mathrm{n},\gamma) {}^{13}\mathrm{B}\)&
1.7$\times$ $10^3$ +9.548$\times$ $10^3$ $T_9$ $^{-3/2}$
$\exp(-1.625/T_9)$&$\times$ 2&
Rauscher et al. 1994\\ 
&&\(+1.562\times 10^3{T_9}^{-3/2}\exp(-2.666/T_9)\)&& \\
&&\(+1.163\times 10^4 {T_9}^{-3/2}\exp(-5.919/T_9)\)&& \\ \hline
(8)&\({}^{13}\mathrm{B}(\mathrm{n},\gamma) {}^{14}\mathrm{B}\)&
\(1.02\times 10^1 + 4.950\times 10^1 T_9\)
&$\times$ 2&Rauscher et al. 1994\\ 
&&\(+4.940\times 10^4{T_9}^{-3/2}\exp(-4.76/T_9)\)&&\\ \hline
(9)&\({}^{14}\mathrm{B}(\mathrm{n},\gamma) {}^{15}\mathrm{B}\)&
$1.906\times 10^{3} + 1.142 \times 10^{3} T_{9}$
&$\times$ 2&\\ \hline
(10)&\({}^{12}\mathrm{C}(\mathrm{n},\gamma) {}^{13}\mathrm{C}\)&
\(4.64\times 10^{2}+5.71\times 10^{3}T_{9}\)
&$\pm$10$\%$&Nagai et al. 1991a\\ \hline
(11)&\({}^{13}\mathrm{C}(\mathrm{n},\gamma) {}^{14}\mathrm{C}\)&
\(1.82\times 10^{2}+4.633\times 10^{4}T_{9}^{-3/2}\exp(-1.636/T_{9})\)
&$\times$ 2&Raman et al. 1990\\ \hline
(12)&\({}^{14}\mathrm{C}(\mathrm{n},\gamma) {}^{15}\mathrm{C}\)&
\(7.8\times 10^{2}T_{9}+2.05\times 10^{3}T_{9}^{-3/2}\exp(-21.14/T_{9})\)
&$\times$ 4&Beer et al. 1992\\ \hline
(13)&\({}^{15}\mathrm{C}(\mathrm{n},\gamma) {}^{16}\mathrm{C}\)&
\(5.27\times 10^{2}T_{9}+3.28\times 10^{4}T_{9}^{-3/2}\exp(-21.56/T_{9})\)
&$\times$ 2&Rauscher et al. 1994\\ \hline
(14)&\({}^{16}\mathrm{C}(\mathrm{n},\gamma) {}^{17}\mathrm{C}\)&
$3.66\times 10^{2}T_{9}$
&$\times$ 2&Rauscher et al. 1994\\ \hline
(15)&\({}^{17}\mathrm{C}(\mathrm{n},\gamma) {}^{18}\mathrm{C}\)&
\(1.100\times 10^3 + 4.05\times 10^1 T_{9} \)
&$\times$ 10&\\
&&\(+ 1.133\times 10^{3}T_{9}^{-3/2}\exp(-0.541/T_{9})\)&&\\ \hline
(16)&\({}^{18}\mathrm{C}(\mathrm{n},\gamma) {}^{19}\mathrm{C}\)&
\(1.014\times 10^3+ 3.377 \times 10^2 T_{9}\)
&$\times$ 2&Nakamura et al. 1999\\ \hline
(17)&\({}^{19}\mathrm{C}(\mathrm{n},\gamma) {}^{20}\mathrm{C}\)&
\(2.10\times 10^2 + 1.02\times 10^1 T_{9}\) 
&$\times$ 10&\\
&&\(+ 3.74\times 10^{2}T_{9}^{-3/2}\exp(-0.75/T_{9})\)&&\\ \hline
(18)&\({}^{18}\mathrm{C}(\alpha,\mathrm{n}) {}^{21}\mathrm{O}\)&
\(1.659\times 10^{13}T_{9}^{-2/3}\exp(-27.5T_{9}^{-1/3})\)
&$\times$ 10&\\ \hline
\label{reaction}
\end{tabular}}\\
a) \(T_{9\ast} = T_{9}/(1.0 + 0.1378T_{9})\),\\
b) Percentage of 1$\sigma$ uncertainty. Otherwise, factor two ($\times$ 2), four ($\times$ 4) and ten ($\times$
 10) uncertainties.\\     
c) Blank means the present estimates, as explained in the text. 
\end{table}

\clearpage


\clearpage

\begin{table}
\centering
\caption{}
\begin{tabular}[t]{ccc}\hline
&Fast Wind Model $^{a}$&Slow Wind Model$^{b}$\\ \hline
$s/\rm{k}$&300&385.7\\ \hline
$\tau_{dyn}$&0.005$\ \rm{s}$&0.300 \rm{s}$^{c}$ \\ \hline
$Y_{e,i}$&0.45&0.3623\\ \hline
$T_{a}$&0.60&1.04$^{c}$\\ \hline
\label{flowmodel}
\end{tabular}
\\
\raggedright
a ~Otsuki et al. (2003).\\
b ~Woosley et al. (1994).\ \ These are the asymptotic values from
 Table 1 of Woosley et al. (1994).\\
c ~Woosley et al. (1994).\ \ These are the values read off from the
 trajectory-40. 
\end{table}

\begin{table}
\centering
\caption{Three exponential-model parameter sets for studying the dependence of the
 r-process on $\tau_{dyn}$. }
\begin{tabular}[t]{ccccc}\hline
$\tau_{dyn}$ &0.50&5.0&50\\ \hline
$s/k$&200&350&1700\\ \hline
$Y_{e}$&0.45&0.45&0.45\\ \hline
$T_{a}$&0.60&0.60&0.60\\ \hline
\label{flowparametertau}
\end{tabular}
\end{table}

\begin{table}
\centering
\caption{The same as table \ref{flowparametertau}, but for $s/k$.}
\begin{tabular}[t]{ccccc}\hline
$\tau_{dyn}$&1.0&1.0&1.0\\ \hline
$s/k$&200&300&400\\ \hline
$Y_{e}$&0.20&0.35&0.45&\\ \hline
$T_{a}$&0.60&0.60&0.60\\ \hline
\label{flowparameterentro}
\end{tabular}
\end{table}

\begin{table}
\centering
\caption{The same as table \ref{flowparametertau}, but for $Y_{e}$ .}
\begin{tabular}[t]{ccccc}\hline
$\tau_{dyn}$&5.0&5.0&5.0\\ \hline
$s/k$&200&300&350\\ \hline
$Y_{e}$&0.20&0.35&0.45&\\ \hline
$T_{a}$&0.60&0.60&0.60\\ \hline
\label{flowparameterye}
\end{tabular}
\end{table}

\begin{table}
\centering
\caption{The same as table \ref{flowparametertau}, but for $T_{a}$.}
\begin{tabular}[t]{ccccc}\hline
$\tau_{dyn}$&5.0&5.0&5.0\\ \hline
$s/k$&250&350&350\\ \hline
$Y_{e}$&0.45&0.45&0.45&\\ \hline
$T_{a}$&0.52&0.62&0.72\\ \hline
\label{flowparametertemp}
\end{tabular}
\end{table}

\begin{table}[p]
\caption{Sensitivity result for the 2nd and 3rd r-process peak elements
 and actinides $^{232}$Th, $^{235}$U and $^{238}$U in the fast flow model.
 The last two column show $\pm$2$\sigma$ current importance of each
 reaction. See the definition in text.}
\centering
\begin{tabular}[t]{rcccccccc}\hline
{No.}&{reaction}&\multicolumn{5}{c}{sensitivity({$\alpha_{i}$})}&\multicolumn{2}{c}{current}\\
{}&{}&{2nd peak}&{3rd
peak}&{$^{232}\rm{Th}$}&{$^{235}\rm{U}$}&{$^{238}\rm{U}$}&\multicolumn{2}{c}{importance($\pm$2$\sigma$)}\\ \hline
(1)&\(\alpha (\alpha \mathrm{n},\gamma ){}^9 \mathrm{Be}
\)&0.1823&-0.6546&-1.9423&-1.9819&-2.1006&0.3445&11.2222
\\ \hline
(2)&\(\alpha(\mathrm{t},\gamma)
{}^7\mathrm{Li}\)&0.2874&-0.7474&-2.7125&-2.7857&-2.9583&0.2658&13.2353
\\ \hline
(3)&\({}^7\mathrm{Li}(\mathrm{n},\gamma)
{}^8\mathrm{Li}\)&0.0465&-0.0917&-0.4296&-0.4436&-0.4729&0.7881& 1.7163
\\ \hline
(4)&\({}^8\mathrm{Li}(\alpha,\mathrm{n})
{}^{11}\mathrm{B}\)&0.0017&-0.0032&-0.0164&-0.0170&-0.0181&0.9882& 1.0120
\\ \hline
(5)&\({}^9\mathrm{Be}(\mathrm{n},\gamma)
{}^{10}\mathrm{Be}\)&0.0042&-0.0105&-0.0337&-0.0346&-0.0365&0.9761& 1.0245
\\ \hline
(6)&\({}^{11}\mathrm{B}(\mathrm{n},\gamma)
{}^{12}\mathrm{B}\)&-0.0100&0.0096&0.1119&0.1166&0.1256&1.0853& 0.9214
\\ \hline
(7)&\({}^{12}\mathrm{B}(\mathrm{n},\gamma)
{}^{13}\mathrm{B}\)&0.0015&-0.0079&-0.0114&-0.0115&-0.0012&0.9944& 1.0056\\ \hline
(8)&\({}^{13}\mathrm{B}(\mathrm{n},\gamma)
{}^{14}\mathrm{B}\)&0.0&0.0&0.0&0.0&0.0&1.0& 1.0\\ \hline
(9)&\({}^{14}\mathrm{B}(\mathrm{n},\gamma)
{}^{15}\mathrm{B}\)&0.00010&-0.0002&-0.0032&-0.0034&-0.0035&0.9977& 1.0024\\ \hline
(10)&\({}^{12}\mathrm{C}(\mathrm{n},\gamma)
{}^{13}\mathrm{C}\)&0.0&0.0&0.0&0.0&0.0&1.0& 1.0\\ \hline
(11)&\({}^{13}\mathrm{C}(\mathrm{n},\gamma)
{}^{14}\mathrm{C}\)&0.0005&-0.0045&-0.0214&-0.0227&-0.0232&0.9846& 1.0157\\ \hline
(12)&\({}^{14}\mathrm{C}(\mathrm{n},\gamma)
{}^{15}\mathrm{C}\)&0.0&0.0&0.0&0.0&0.0&1.0& 1.0\\ \hline
(13)&\({}^{15}\mathrm{C}(\mathrm{n},\gamma)
{}^{16}\mathrm{C}\)&0.0040&-0.0194&-0.0899&-0.0878&-0.0867&0.9407& 1.0630\\ \hline
(14)&\({}^{16}\mathrm{C}(\mathrm{n},\gamma)
{}^{17}\mathrm{C}\)&0.0&0.0&0.0&0.0&0.0&1.0& 1.0\\ \hline
(15)&\({}^{17}\mathrm{C}(\mathrm{n},\gamma)
{}^{18}\mathrm{C}\)&0.0274&-0.0209&-0.1624&-0.1735&-0.1767&0.6747& 1.4821
\\ \hline
(16)&\({}^{18}\mathrm{C}(\mathrm{n},\gamma)
{}^{19}\mathrm{C}\)&0.0&0.0&0.0&0.0&0.0&1.0& 1.0\\ \hline
(17)&\({}^{19}\mathrm{C}(\mathrm{n},\gamma)
{}^{20}\mathrm{C}\)&0.0&0.0&0.0&0.0&0.0&1.0& 1.0\\ \hline
(18)&\({}^{18}\mathrm{C}(\alpha,\mathrm{n})
{}^{21}\mathrm{O}\)&0.0233&-0.0017&-0.0285&-0.0288&-0.0298&0.9354& 1.0691\\ \hline
\label{result1}
\end{tabular}
\end{table}

\begin{table}
\caption{Sensitivity result for the 2nd r-process peak elements in the
 slow wind model. The last two column show $\pm$2$\sigma$ current importance of each
 reaction. See the definition in text.}
\centering
\begin{tabular}[t]{rcccc}\hline
{No.}&{reaction}&{sensitivity({$\alpha_{i}$})}&\multicolumn{2}{c}{current
 importance($\pm$2$\sigma$)}\\ \hline
(1)&\(\alpha (\alpha \mathrm{n},\gamma ){}^9 \mathrm{Be} \)
&0.2388&1.1351& 0.7502
\\ \hline
(2)&\(\alpha(\mathrm{t},\gamma) {}^7\mathrm{Li}\)
&-0.1377&0.9373& 1.1345
\\ \hline
(3)&\({}^7\mathrm{Li}(\mathrm{n},\gamma) {}^8\mathrm{Li}\)
&-0.0486&0.9745& 1.0603
\\ \hline
(4)&\({}^8\mathrm{Li}(\alpha,\mathrm{n}) {}^{11}\mathrm{B}\)
&-0.0056&0.9961& 1.0039
\\ \hline
(5)&\({}^9\mathrm{Be}(\mathrm{n},\gamma) {}^{10}\mathrm{Be}\)
&0.0&1.0&1.0
\\ \hline
(6)&\({}^{11}\mathrm{B}(\mathrm{n},\gamma) {}^{12}\mathrm{B}\)
&-0.0005&0.9997& 1.0003
\\ \hline
(7)&\({}^{12}\mathrm{B}(\mathrm{n},\gamma) {}^{13}\mathrm{B}\)
&0.0&1.0&1.0
\\ \hline
(8)&\({}^{13}\mathrm{B}(\mathrm{n},\gamma) {}^{14}\mathrm{B}\)
&0.0&1.0&1.0
\\ \hline
(9)&\({}^{14}\mathrm{B}(\mathrm{n},\gamma) {}^{15}\mathrm{B}\)
&0.0&1.0&1.0
\\ \hline
(10)&\({}^{12}\mathrm{C}(\mathrm{n},\gamma) {}^{13}\mathrm{C}\)
&0.0&1.0&1.0
\\ \hline
(11)&\({}^{13}\mathrm{C}(\mathrm{n},\gamma) {}^{14}\mathrm{C}\)
&-0.0016&0.9963& 1.0037
\\ \hline
(12)&\({}^{14}\mathrm{C}(\mathrm{n},\gamma) {}^{15}\mathrm{C}\)
&0.0&1.0&1.0
\\ \hline
(13)&\({}^{15}\mathrm{C}(\mathrm{n},\gamma) {}^{16}\mathrm{C}\)
&0.0&1.0&1.0
\\ \hline
(14)&\({}^{16}\mathrm{C}(\mathrm{n},\gamma) {}^{17}\mathrm{C}\)
&0.0&1.0&1.0
\\ \hline
(15)&\({}^{17}\mathrm{C}(\mathrm{n},\gamma) {}^{18}\mathrm{C}\)
&0.0&1.0&1.0
\\ \hline
(16)&\({}^{18}\mathrm{C}(\mathrm{n},\gamma) {}^{19}\mathrm{C}\)
&0.0&1.0&1.0
\\ \hline
(17)&\({}^{19}\mathrm{C}(\mathrm{n},\gamma) {}^{20}\mathrm{C}\)
&0.0&1.0&1.0
\\ \hline
(18)&\({}^{18}\mathrm{C}(\alpha,\mathrm{n}) {}^{21}\mathrm{O}\)
&0.0&1.0&1.0
\\ \hline
\label{result2}
\end{tabular}
\end{table}

\clearpage 

\begin{table}[p]
\caption{Sensitivity results for the 2nd and 3rd
 r-process peak elements and actinides  Th, $^{235}\mathrm{U}$ and $^{238}\mathrm{U}$. The parameters
 of the fast wind model are ${\tau}_{dyn}$=0.5 ms, s/k=200 and
 $Y_{\mathrm{e}}$=0.45.  The last two column show $\pm$2$\sigma$ current importance of each
 reaction. See the definition in text.}
\begin{tabular}[t]{rccccccrl}\hline
{No.}&{reaction}&{2nd peak}&{3rd
peak}&{Th}&{$^{235}\mathrm{U}$}&{$^{238}\mathrm{U}$}&\multicolumn{2}{c}{current
 importance($\pm$2$\sigma$)}
\\ \hline
(1)&\(\alpha (\alpha \mathrm{n},\gamma ){}^9 \mathrm{Be} \)&0.0632& -0.6108&
-1.4546& -1.4689& -1.5412&0.4540& 6.0002
\\ \hline
(2)&\(\alpha(\mathrm{t},\gamma) {}^7\mathrm{Li}\)&0.0498& -0.9534& -2.1597&
-2.1774& -2.2954&0.3538& 7.5819
\\ \hline
(3)&\({}^7\mathrm{Li}(\mathrm{n},\gamma) {}^8\mathrm{Li}\)&-0.0050&  0.0197&
0.0641&  0.0641&  0.0657&1.0349& 0.9251
\\ \hline
(4)&\({}^8\mathrm{Li}(\alpha,\mathrm{n}) {}^{11}\mathrm{B}\)&-0.0004&
0.0012&  0.0047&  0.0048&  0.0048&1.0033& 0.9967
\\ \hline
(5)&\({}^9\mathrm{Be}(\mathrm{n},\gamma) {}^{10}\mathrm{Be}\)&0.0001&
-0.0009& -0.0027& -0.0027& -0.0029&0.9981& 1.0019
\\ \hline
(6)&\({}^{11}\mathrm{B}(\mathrm{n},\gamma) {}^{12}\mathrm{B}\)&0.0015&
-0.0073& -0.0200& -0.0199& -0.0209&0.9861& 1.0141
\\ \hline
(15)&\({}^{17}\mathrm{C}(\mathrm{n},\gamma) {}^{18}\mathrm{C}\)&-0.0032&
-0.0169& -0.0402& -0.0405& -0.0442&0.9086& 1.1006
\\ \hline
\label{resulta}
\end{tabular}
\end{table}

\begin{table}[p]
\caption{The same as table \ref{resulta}, but for 
the parameters of the fast wind model ${\tau}_{dyn}$=5 ms, s/k=350 and $Y_{\mathrm{e}}$=0.45.}
\begin{tabular}[t]{rccccccrl}\hline
{No.}&{reaction}&{2nd peak}&{3rd
peak}&{Th}&{$^{235}\mathrm{U}$}&{$^{238}\mathrm{U}$}&\multicolumn{2}{c}{current
 importance($\pm$2$\sigma$)}
\\ \hline
(1)&\(\alpha (\alpha \mathrm{n},\gamma ){}^9 \mathrm{Be} \)&0.0906& -0.2290&
-0.7554& -0.7661& -0.8133&0.6617&2.5523
\\ \hline
(2)&\(\alpha(\mathrm{t},\gamma) {}^7\mathrm{Li}\)&0.6500& -1.2696& -4.9599&
-5.0445& -5.3809&0.0898&109.8522
\\ \hline
(3)&\({}^7\mathrm{Li}(\mathrm{n},\gamma) {}^8\mathrm{Li}\)&0.1244& -0.1587&
-0.8138& -0.8305& -0.8879&0.6390&2.7628
\\ \hline
(4)&\({}^8\mathrm{Li}(\alpha,\mathrm{n}) {}^{11}\mathrm{B}\)&0.0046&
-0.0051& -0.0298& -0.0305& -0.0325&0.9788&1.0271
\\ \hline
(5)&\({}^9\mathrm{Be}(\mathrm{n},\gamma) {}^{10}\mathrm{Be}\)&0.0025&
-0.0043& -0.0158& -0.0160& -0.0171&0.9888&1.0114
\\ \hline
(6)&\({}^{11}\mathrm{B}(\mathrm{n},\gamma) {}^{12}\mathrm{B}\)&-0.0049&
-0.0171&  0.0583&  0.0599&  0.0690&1.0442&0.9577
\\ \hline
(15)&\({}^{17}\mathrm{C}(\mathrm{n},\gamma) {}^{18}\mathrm{C}\)&0.0124&
-0.0356& -0.1271& -0.1278& -0.1409&0.7380&1.3550
\\ \hline
\label{resultb}
\end{tabular}
\end{table}

\begin{table}[p]
\caption{The same as table \ref{resulta}, but for the 
 parameters of the fast wind model 
 ${\tau}_{dyn}$=50 ms, s/k=1700 and $Y_{\mathrm{e}}$=0.45.}
\begin{tabular}[t]{rccccccrl}\hline
{No.}&{reaction}&{2nd peak}&{3rd
peak}&{Th}&{$^{235}\mathrm{U}$}&{$^{238}\mathrm{U}$}&\multicolumn{2}{c}{current
 importance($\pm$2$\sigma$)}
\\ \hline
(1)&\(\alpha (\alpha \mathrm{n},\gamma ){}^9 \mathrm{Be} \)&0.0045& -0.0099&
-0.0304& -0.0306& -0.0330&0.9835&1.0384
\\ \hline
(2)&\(\alpha(\mathrm{t},\gamma) {}^7\mathrm{Li}\)&1.5255& -1.9525& -8.4970&
-8.6190& -9.1254&0.0164&3025.7589
\\ \hline
(3)&\({}^7\mathrm{Li}(\mathrm{n},\gamma) {}^8\mathrm{Li}\)&0.5260& -0.5790&
-2.8761& -2.9151& -3.0879&0.2079&35.2829
\\ \hline
(4)&\({}^8\mathrm{Li}(\alpha,\mathrm{n}) {}^{11}\mathrm{B}\)&0.0552&
-0.0613& -0.2939& -0.2985& -0.3155&0.8108&1.2334
\\ \hline
(5)&\({}^9\mathrm{Be}(\mathrm{n},\gamma) {}^{10}\mathrm{Be}\)&0.0005&
-0.0012& -0.0029& -0.0028& -0.0031&0.9980&1.0020
\\ \hline
(6)&\({}^{11}\mathrm{B}(\mathrm{n},\gamma) {}^{12}\mathrm{B}\)&0.0009&
-0.0294& -0.0295& -0.0295& -0.0282&0.9801&1.0204
\\ \hline
(15)&\({}^{17}\mathrm{C}(\mathrm{n},\gamma) {}^{18}\mathrm{C}\)&0.0779&
-0.1152& -0.4049& -0.4111& -0.4347&0.3829&2.6116
\\ \hline
\label{resultc}
\end{tabular}
\end{table}

\begin{table}[p]
\caption{The same as table \ref{resulta}, but for the 
 parameters of the fast wind 
 model ${\tau}_{dyn}$=5 ms, s/k=200 and $Y_{\mathrm{e}}$=0.20}
\begin{tabular}[t]{rccccccrl}\hline
{No.}&{reaction}&{2nd peak}&{3rd
peak}&{Th}&{$^{235}\mathrm{U}$}&{$^{238}\mathrm{U}$}&\multicolumn{2}{c}{current
 importance($\pm$2$\sigma$)}
\\ \hline
(1)&\(\alpha (\alpha \mathrm{n},\gamma ){}^9 \mathrm{Be} \)&0.0767& -0.1616&
-0.2529& -0.2416& -0.2585&0.8753& 1.3528
\\ \hline
(2)&\(\alpha(\mathrm{t},\gamma) {}^7\mathrm{Li}\)&0.8267& -1.2469& -2.5047&
-2.5057& -2.4766&0.3094& 9.8430
\\ \hline
(3)&\({}^7\mathrm{Li}(\mathrm{n},\gamma) {}^8\mathrm{Li}\)&0.1484& -0.1782&
-0.4368& -0.4371& -0.4415&0.7924& 1.6954
\\ \hline
(4)&\({}^8\mathrm{Li}(\alpha,\mathrm{n}) {}^{11}\mathrm{B}\)&0.0283&
-0.0335& -0.0812& -0.0810& -0.0821&0.9451& 1.0581
\\ \hline
(5)&\({}^9\mathrm{Be}(\mathrm{n},\gamma) {}^{10}\mathrm{Be}\)&0.0084&
-0.0112& -0.0235& -0.0236& -0.0231&0.9839& 1.0164
\\ \hline
(6)&\({}^{11}\mathrm{B}(\mathrm{n},\gamma) {}^{12}\mathrm{B}\)&-0.0247&
-0.0072&  0.1097&  0.1115&  0.1181&1.0815& 0.9246
\\ \hline
(15)&\({}^{17}\mathrm{C}(\mathrm{n},\gamma) {}^{18}\mathrm{C}\)&-0.0025&
0.0102& -0.0022& -0.0028& -0.0030&0.9939& 1.0062
\\ \hline
\label{resulte}
\end{tabular}
\end{table}

\begin{table}[p]
\caption{The same as table \ref{resulta}, but for the 
 parameters of the fast wind 
 model ${\tau}_{dyn}$=5 ms, s/k=300 and $Y_{\mathrm{e}}$=0.35.}
\begin{tabular}[t]{rccccccrl}\hline
{No.}&{reaction}&{2nd peak}&{3rd
peak}&{Th}&{$^{235}\mathrm{U}$}&{$^{238}\mathrm{U}$}&\multicolumn{2}{c}{current
 importance($\pm$2$\sigma$)}
\\ \hline
(1)&\(\alpha (\alpha \mathrm{n},\gamma ){}^9 \mathrm{Be} \)&0.0879& -0.1810&
-0.5383& -0.5502& -0.5724&0.7454&1.9475
\\ \hline
(2)&\(\alpha(\mathrm{t},\gamma) {}^7\mathrm{Li}\)&0.7862& -1.2755& -4.4828&
-4.5748& -4.8373&0.1134&69.6807
\\ \hline
(3)&\({}^7\mathrm{Li}(\mathrm{n},\gamma) {}^8\mathrm{Li}\)&0.1466& -0.1774&
-0.7737& -0.7914& -0.8407&0.6534&2.6261
\\ \hline
(4)&\({}^8\mathrm{Li}(\alpha,\mathrm{n}) {}^{11}\mathrm{B}\)&0.0118&
-0.0102& -0.0552& -0.0580& -0.0577&0.9613&1.0403
\\ \hline
(5)&\({}^9\mathrm{Be}(\mathrm{n},\gamma) {}^{10}\mathrm{Be}\)&0.0059&
-0.0097& -0.0318& -0.0320& -0.0349&0.9775&1.0231
\\ \hline
(6)&\({}^{11}\mathrm{B}(\mathrm{n},\gamma) {}^{12}\mathrm{B}\)&-0.0222&
-0.0078&  0.1662&  0.1714&  0.1887&1.1293& 0.8855
\\ \hline
(15)&\({}^{17}\mathrm{C}(\mathrm{n},\gamma) {}^{18}\mathrm{C}\)&0.0021&
0.0021& -0.0268& -0.0275& -0.0307&0.9368& 1.0674
\\ \hline
\label{resultd}
\end{tabular}
\end{table}

\begin{table}
\caption{$\beta$-decay lifetimes in units of s, and the neutron
 separation energies ($S_{n}$) in units of MeV.
 The $S_{\rm{n}}$-value for $^{19} \rm{C}$ was changed from 0.191 MeV to
 0.530 MeV (Nakamura et al. 1999).}
\begin{tabular}[t]{ccccccccccccc}\hline
{}&\multicolumn{2}{c}{\({}^{15}\mathrm{C}\)}&\multicolumn{2}{c}{\({}^{16}\mathrm{C}\)}&
\multicolumn{2}{c}{\({}^{17}\mathrm{C}\)}&\multicolumn{2}{c}{\({}^{18}\mathrm{C}\)}&
\multicolumn{2}{c}{\({}^{19}\mathrm{C}\)}&\multicolumn{2}{c}{\({}^{20}\mathrm{C}\)}\\
\hline
{$\beta$-life time[s]}&\multicolumn{2}{c}{2.449}&
\multicolumn{2}{c}{0.747}&\multicolumn{2}{c}{0.193}&
\multicolumn{2}{c}{0.092}&\multicolumn{2}{c}{0.049}&\multicolumn{2}{c}{0.014}\\
\hline
{$\mathrm{S_n}$ [$\mathbf{MeV}$]}&{}&\multicolumn{2}{c}{4.250}&
\multicolumn{2}{c}{0.725}&\multicolumn{2}{c}{4.191}&
\multicolumn{2}{c}{0.191$\rightarrow$0.530}&\multicolumn{2}{c}{3.345}&{}\\
 \hline\\
\end{tabular}
\label{tablecarbon}
\end{table}
\end{document}